\documentclass[fleqn,usenatbib]{mnras}

\usepackage{newtxtext,newtxmath}

\usepackage[T1]{fontenc}
\usepackage{ae,aecompl}

\usepackage{graphicx}	
\usepackage{amsmath}	
\usepackage{amssymb}	
\usepackage[flushleft]{threeparttable}

\usepackage{xcolor}
\usepackage{url}						

\definecolor{crimson}{HTML}{DC143C}
\definecolor{darkred}{HTML}{8B0000}

\hypersetup{citecolor=purple}

\defcitealias{leisman}{L17}

\title[Gas kinematics and formation of gas-rich UDGs]{Robust H\,{\sc i} kinematics of gas-rich ultra-diffuse galaxies: hints of a weak-feedback formation scenario}

\author[Pavel E. Mancera Pi\~na et al.]{Pavel E. Mancera Pi\~na$^{1,2}$\thanks{e-mail: pavel@astro.rug.nl},
Filippo Fraternali$^{1}$,  Kyle A. Oman$^{1,3}$, \newauthor Elizabeth A. K. Adams$^{2,1}$, Cecilia Bacchini$^{1,4,5}$, Antonino Marasco$^{6}$, Tom Oosterloo$^{2,1}$, \newauthor Gabriele Pezzulli$^{7}$, Lorenzo Posti$^{8}$, Lukas Leisman$^{9}$, John M. Cannon$^{10}$, \newauthor Enrico M. di Teodoro$^{11}$, Lexi Gault$^{9}$, Martha P. Haynes$^{12}$, Kameron Reiter$^{9}$,\newauthor Katherine L. Rhode$^{13}$, John J. Salzer$^{13}$ and Nicholas J. Smith$^{13}$\\
\\
$^{1}$ Kapteyn Astronomical Institute, University of Groningen, Landleven 12, 9747 AD, Groningen, The Netherlands\\
$^{2}$ ASTRON, Netherlands Institute for Radio Astronomy, Postbus 2, 7900 AA Dwingeloo, The Netherlands\\
$^{3}$  Institute for Computational Cosmology, Department of Physics, Durham University, Science Laboratories, South Road, Durham DH1 3LE, UK\\
$^{4}$ Dipartimento di Fisica e Astronomia, Università di Bologna, via Gobetti 93/2, I-40129, Bologna, Italy\\
$^{5}$ INAF--Osservatorio di Astrofisica e Scienza dello Spazio di Bologna, via Gobetti 93/3, I-40129 Bologna, Italy\\
$^{6}$ INAF--Osservatorio Astrofisico di Arcetri, Largo E. Fermi 5, I-50157, Firenze, Italy\\
$^{7}$ Department of Physics, ETH Zurich, Wolfgang-Pauli-Strasse 27, CH-8093 Zurich, Switzerland\\
$^{8}$ Universit\'e de Strasbourg, CNRS UMR 7550, Observatoire astronomique de Strasbourg, 11 rue de l'Universit\'e, 67000 Strasbourg, France\\
$^{9}$ Department of Physics and Astronomy, Valparaiso University, Neils Science Center, 1610 Campus Drive East, Valparaiso, IN 46383, USA\\
$^{10}$ Department of Physics \& Astronomy, Macalester College, 1600 Grand Avenue, Saint Paul, MN 55105, USA\\
$^{11}$ Department of Physics \& Astronomy, Johns Hopkins University, Baltimore, MD 21218, USA \\
$^{12}$ Cornell Center for Astrophysics and Planetary Science, Space  Sciences Building, Cornell University, Ithaca,  NY 14853, USA\\
$^{13}$ Department of Astronomy, Indiana University, 727 East Third Street, Bloomington, IN 47405, USA}
\date{}

\pubyear{2019} 

\usepackage{etoolbox}
\makeatletter
\patchcmd\@combinedblfloats{\box\@outputbox}{\unvbox\@outputbox}{}{\errmessage{\noexpand patch failed}}
\makeatother

\begin{document}
\label{firstpage}
\pagerange{\pageref{firstpage}--\pageref{lastpage}}
\maketitle

\begin{abstract}
We study the gas kinematics of a sample of six isolated gas-rich low surface brightness galaxies, of the class called ultra-diffuse galaxies (UDGs). These galaxies have recently been shown to be outliers from the baryonic Tully-Fisher relation (BTFR), as they rotate much slower than expected given their baryonic mass, and to have baryon fractions similar to the cosmological mean. 
By means of a 3D kinematic modelling fitting technique, we show that the H\,{\sc i} in our UDGs is distributed in `thin' regularly rotating discs and we determine their rotation velocity and gas velocity dispersion. We revisit the BTFR adding galaxies from other studies. We find a previously unknown trend between the deviation from the BTFR and the disc scale length valid for dwarf galaxies with circular speeds $\lesssim$~45~km~s$^{-1}$, with our UDGs being at the extreme end.
Based on our findings, we suggest that the high baryon fractions of our UDGs may originate due to the fact that they have experienced weak stellar feedback, likely due to their low star formation rate surface densities, and as a result they did not eject significant amounts of gas out of their discs. At the same time, we find indications that our UDGs may have higher-than-average stellar specific angular momentum, which can explain their large optical scale lengths.

\end{abstract}



\section{Introduction}
In the last five years there have been a significant number of studies aiming to detect and systematically characterize a population of low surface brightness (LSB) galaxies with Milky Way-like effective radius, similar to those earlier reported by \citet{sandage} or \citet{impey}. Following the work by \citet{vandokkum}, who discovered 47 of these so-called ultra-diffuse galaxies (UDGs), different studies have found them in both high- and low-density environments (e.g., \citealt{vanderburg, roman, roman2, greco, paperII, roman3}, and references therein). Among them, \citet{leisman}, hereafter \citetalias{leisman}, found a population of field galaxies, detected in the ALFALFA catalogue \citep{alfalfa}, that meet the usual optical definition for UDGs ($\langle \upmu(r,R_{\rm e})\rangle \gtrsim$~24~mag~arcsec$^{-2}$, $R_{\rm e} \gtrsim$~1.5~kpc\footnote{With $\langle \upmu(r,R_{\rm e})\rangle$ the mean effective surface brightness within the effective radius, measured in the $r-$band, and $R_{\rm e}$ the optical effective (half-light) radius.}), but have also large atomic gas reservoirs ($\geq 10^8$~M$_\odot$), in contrast to the cluster population.
This gas-rich field population is likely to be small in terms of total number. \citet{prole} estimated that gas-rich UDGs represent about one-fifth of the overall UDG population (cf. \citealt{paperI,lee2020}), and \citet{jones} found that they represent a small correction to the galaxy stellar and H\,{\sc i} mass functions at all masses, with a maximum contribution to the H\,{\sc i} mass function of 6 per cent at $\sim$10$^{9}$~M$_\odot$. Despite this, their extreme properties make them puzzling and interesting objects to study. 

It is well known that resolved 21-cm observations not only reveal interactions between the extended H\,{\sc i} galaxy discs and their environments (e.g., \citealt{yun,deblok2000,ff02,oosterloo2007,enrico14}), but also allow us to estimate their rotation velocity, angular momentum and matter distribution, key ingredients to understand their formation and evolution (e.g., \citealt{deblok,marc,swaters,noordermeer,posti2}). Because of these key properties, that may reveal telltale clues about their origins, pursuing studies of UDGs from an H\,{\sc i} perspective is potentially very interesting.

From a theoretical perspective, different ideas have been proposed to explain the puzzling nature of UDGs. \citet{dicintio} presented hydrodynamical simulations where UDGs originate in isolation due to powerful feedback-driven outflows that modify the dark matter density profile allowing the baryons to move to external orbits, increasing the scale length of the galaxies (see also \citealt{chan,nihaoxxiv}). On the other hand, \citet{amorisco} suggested that the extended sizes of UDGs can be explained if they live in dark matter haloes with high spin parameter (see also \citealt{rong,posti}). While currently those seem to be the most popular ideas, more mechanisms have been proposed in the literature, as we discuss in detail later.

To test these theories, \textit{isolated} UDGs are very useful. Some of their properties like morphology, circular speed, baryon fraction or angular momentum, can be contrasted with expectations from the above mentioned theories in a relatively straightforward way, since they are not affected by their environments and cannot be explained by interactions with other galaxies (e.g., \citealt{Aku,bennet}). Using a combination of H\,{\sc i} interferometric data and deep optical images for a sample of six gas-rich UDGs, \citet{btfr} studied the baryonic mass--circular speed plane, finding that these galaxies show a set of intriguing properties: they lie well above the canonical baryonic Tully-Fisher relation (BTFR, \citealt{mcgaugh1}), in a position compatible with having `no missing baryons' within their virial radii, and with little room for dark matter inside the extent of their gaseous discs. 

In this work we delve into the kinematic properties of the galaxies presented in \citet{btfr}, explaining in detail the methodology used to derive 3D kinematic models. Further, we expand our investigation to other properties of these LSB galaxies, and discuss possible interpretations for our results.

The rest of this manuscript is organised as follows. In Section~\ref{sec:data} we describe our sample and give the structural parameters obtained from the optical and H\,{\sc i} observations, and in Section~\ref{sec:kinematics} we provide details on our methodology and kinematic modelling. In Section~\ref{sec:ISM} we estimate the scale height of the sample and we look briefly into the properties of their interstellar medium (ISM), while in Section~\ref{sec:btfr} we revisit the BTFR, examine the existence of outliers and show that the deviation from the relation at low rotation velocities correlates with the galaxy scale length. A discussion on the implications of our results for proposed UDG formation mechanisms, including the addition of UDGs to the stellar specific angular momentum--mass relation, is given in Section~\ref{sec:origins}. In Section~\ref{sec:conclusions} we present our conclusions.

\noindent
Throughout this work magnitudes are in the AB system, and a $\Lambda$CDM cosmology with $\Omega_\textnormal{m} = 0.3$, $\Omega_{\Lambda} = 0.7$ and $\mathrm{H_0} = 70$~km~s$^{-1}$~Mpc$^{-1}$ is adopted.

\section{The sample}
\label{sec:data}
The sample studied in this work and in \citet{btfr} consists of six gas-rich UDGs, originally identified by \citetalias{leisman}, for which dedicated optical and interferometric observations were obtained. The observations and data reduction strategies are explained in detail in \citetalias{leisman} and \citet{gault}. We note here that the sample from \citet{gault} consists of eleven galaxies while ours consists of six. As briefly discussed in \citet{btfr}, we selected the galaxies that were more suitable in terms of data-quality for our kinematic modelling (see below). Figures~\ref{fig:114905}--\ref{fig:749290} present our data and 3D kinematic modelling, while Table~\ref{tab:properties} gives the main properties of our galaxy sample.

\begin{figure*}
    \centering
    \includegraphics[scale=00.51]{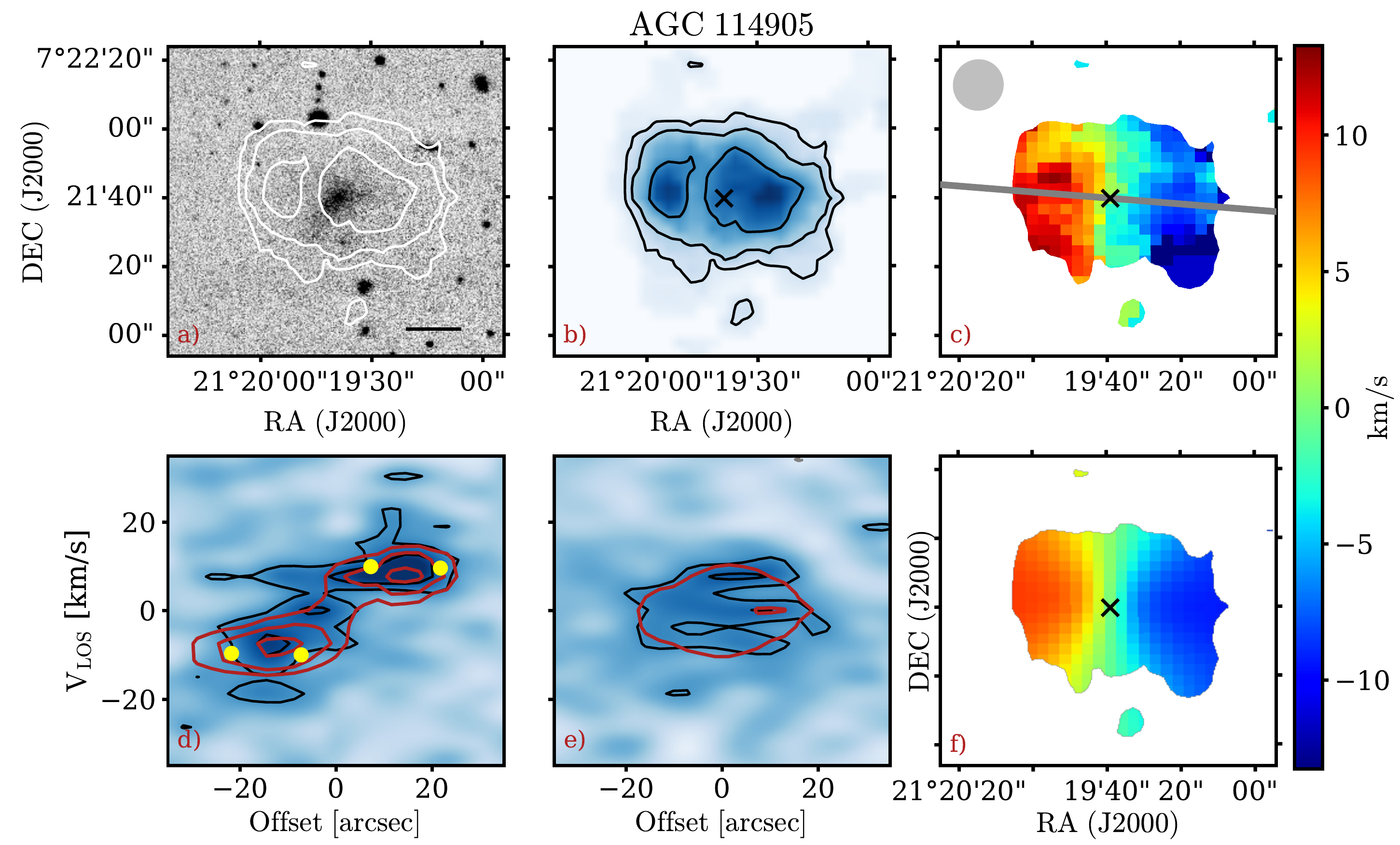}
    \caption{Data and kinematic models for the gas-rich UDG AGC~114905. \textit{a)}: $r-$ band image with H\,{\sc i} contours on top at 1, 2, 4$\times$10$^{20}$ atoms~cm$^{-2}$, with the lowest one at S/N~$\approx$~3. The black solid line indicates a physical scale of 5~kpc. \textit{b)}: total H\,{\sc i} map in blue, and contours as in panel \textit{a)}. \textit{c)}: Observed velocity field (first-moment map). The grey line shows the major axis, while the grey ellipse shows the beam. \textit{d)}: PV diagram along the major axis. Black and red contours correspond to data and best-fit model, respectively, and are at the 2$\upsigma$ and 4$\upsigma$ levels. If present, grey dashed contours indicate negative values in the data. The recovered rotation velocities are indicated with the yellow points. \textit{e)}: PV diagram along the minor axis (perpendicular to the major axis), colours as in \textit{d)}. \textit{f)}: Modelled velocity field. The black cross in panel \textit{b}, \textit{c} and \textit{f} shows the kinematic centre. The rightmost panel shows the velocity colour-bar for panels \textit{c)} and \textit{f)}.}
    \label{fig:114905}
\end{figure*}{}

\begin{figure*}
    \centering
    \includegraphics[scale=00.51]{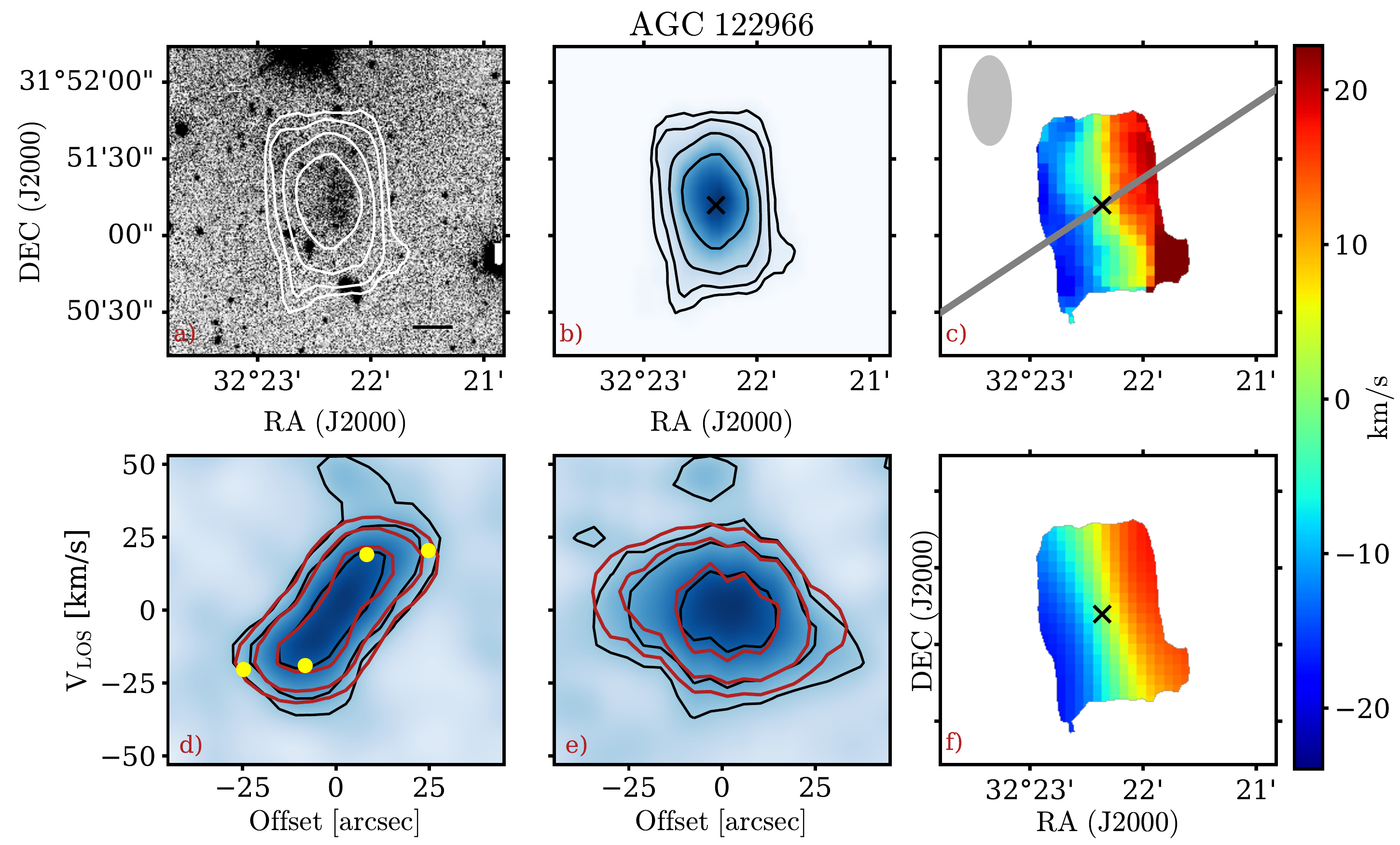}
    \caption{Data and kinematic models for the gas-rich UDG AGC~122966. Panels and symbols as in Figure~\ref{fig:114905}. The H\,{\sc i} contours are at 0.35, 0.7, 1.4 and 2.8$\times$10$^{20}$ atoms~cm$^{-2}$. Note that the kinematic and morphological position angles seem to be different, but this apparent effect is due to the peculiar elongated shape of the WSRT beam (see Appendix~\ref{app:extra}}.
    \label{fig:122966}
\end{figure*}{}

\begin{figure*}
    \centering
    \includegraphics[scale=00.51]{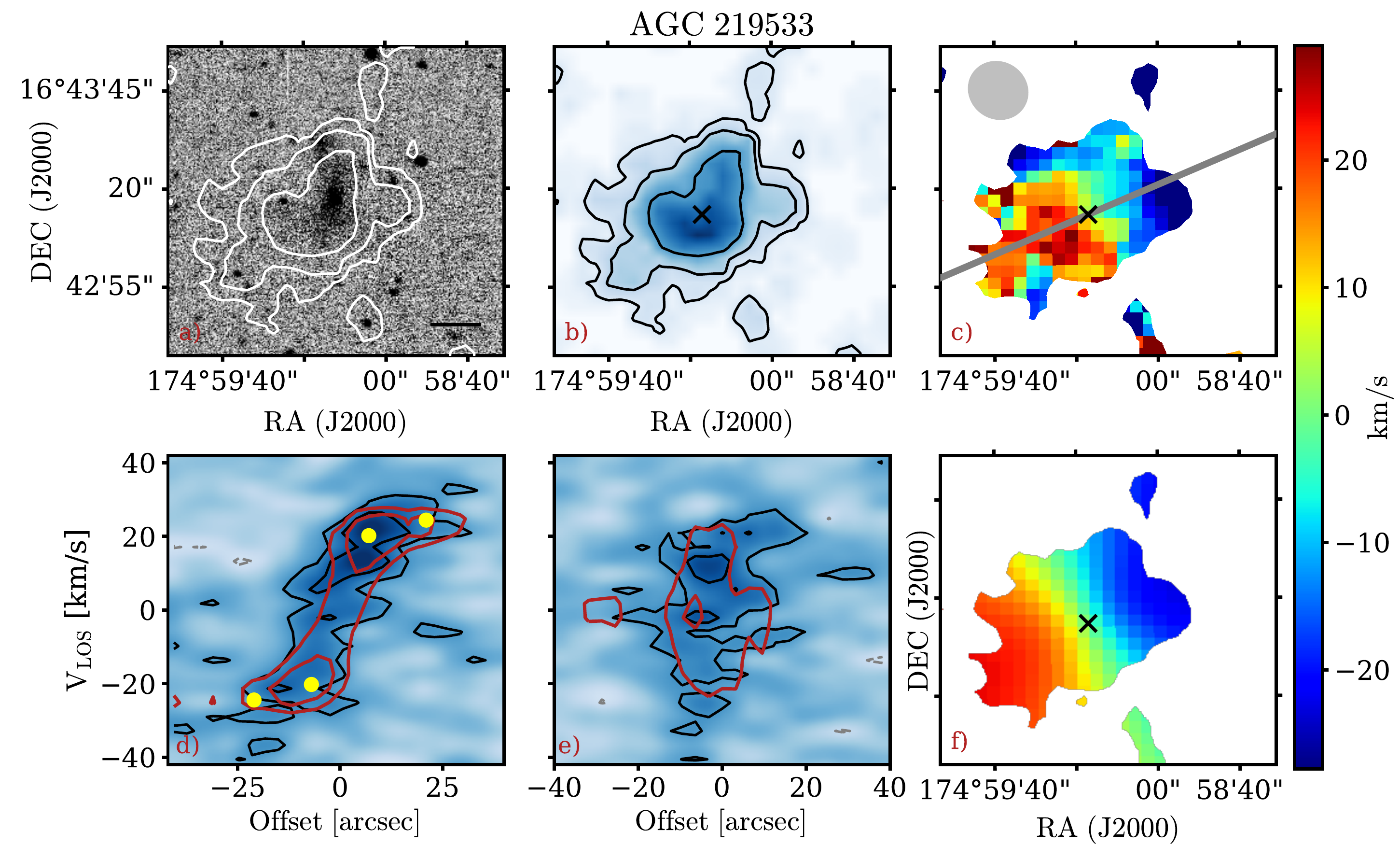}
    \caption{Data and kinematic models for the gas-rich UDG AGC~219533. Panels and symbols as in Figure~\ref{fig:114905}. The H\,{\sc i} contours are at 1.1, 2.2 and 4.4$\times$10$^{20}$ atoms~cm$^{-2}$. In this case the data cube is more noisy than in the rest of the sample, and the last contour corresponds to S/N~$\approx$ 5.}
    \label{fig:219533}
\end{figure*}{}

\begin{figure*}
    \centering
    \includegraphics[scale=00.51]{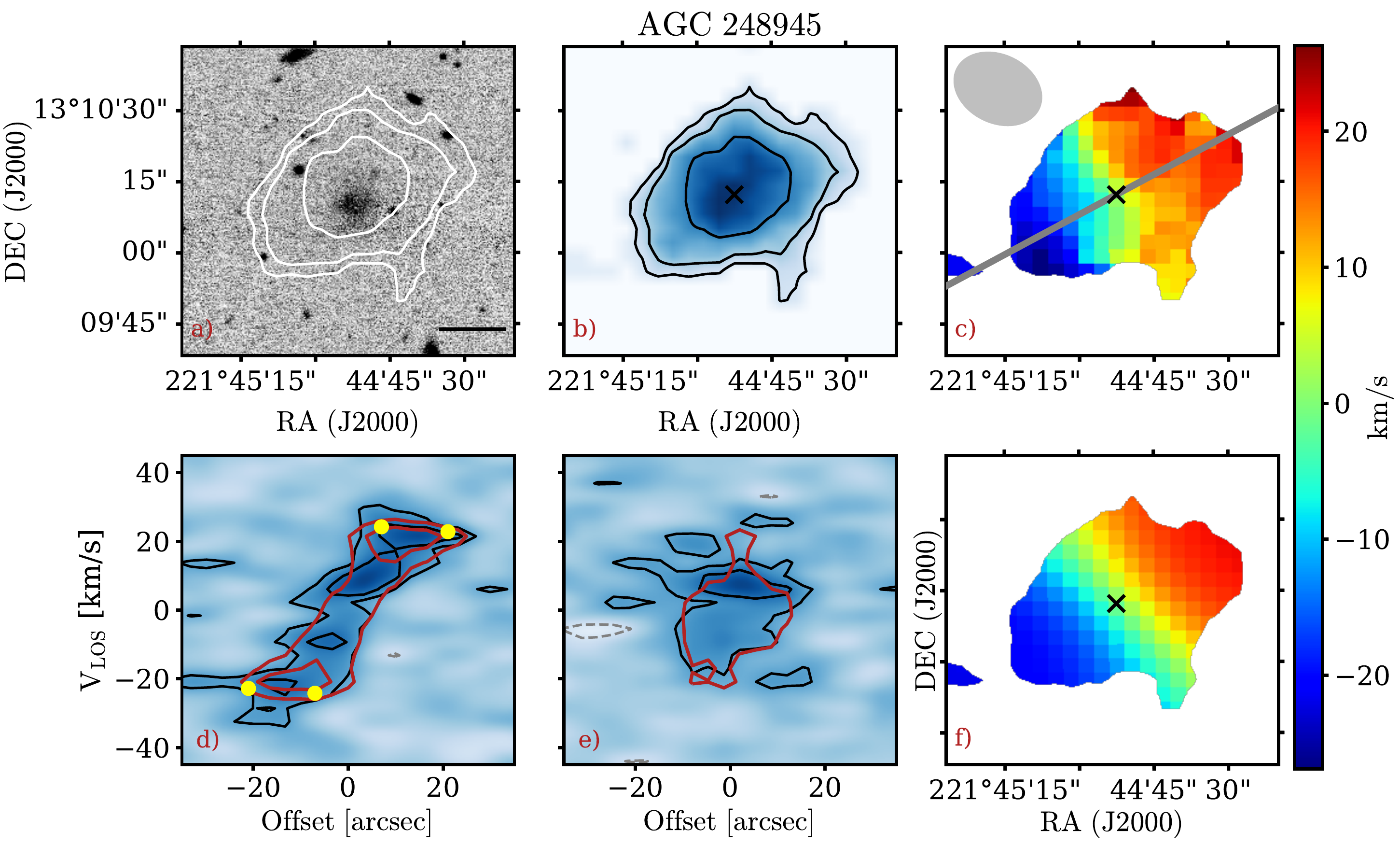}
    \caption{Data and kinematic models for the gas-rich UDG AGC~248945. Panels and symbols as in Figure~\ref{fig:114905}. The H\,{\sc i} contours are at 0.8, 1.6 and 3.2$\times$10$^{20}$ atoms~cm$^{-2}$.}
    \label{fig:248945}
\end{figure*}{}

\begin{figure*}
    \centering
    \includegraphics[scale=00.51]{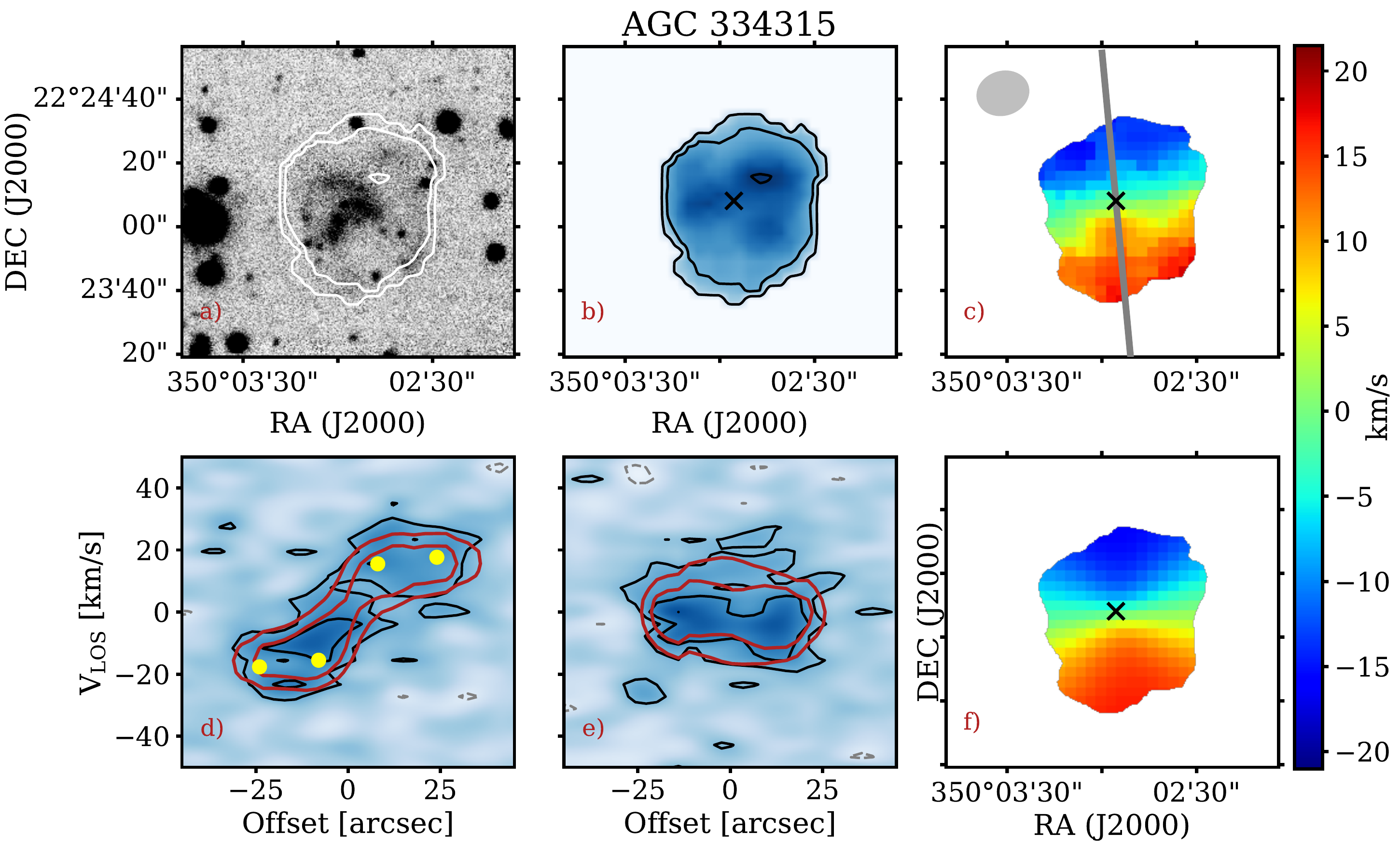}
    \caption{Data and kinematic models for the gas-rich UDG AGC~334315. Panels and symbols as in Figure~\ref{fig:114905}. The H\,{\sc i} contours are at 1.8, 3.6 and 7.2$\times$10$^{20}$ atoms~cm$^{-2}$.}
    \label{fig:334315}
\end{figure*}{}

\begin{figure*}
    \centering
    \includegraphics[scale=00.51]{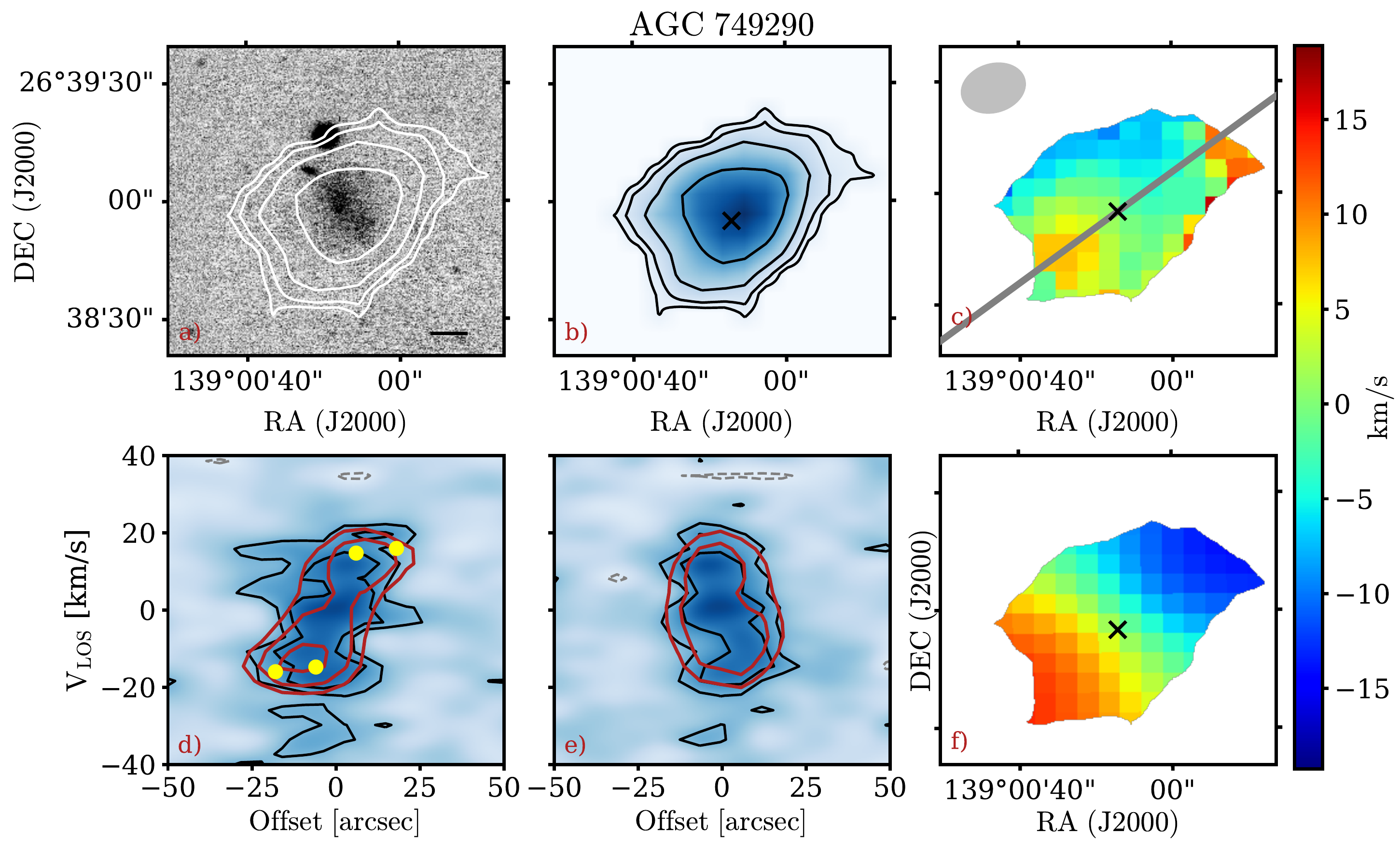}
    \caption{Data and kinematic models for the gas-rich UDG AGC~749290. Panels and symbols as in Figure~\ref{fig:114905}. The H\,{\sc i} contours are at 0.35, 0.7, 1.4 and 2.8$\times$10$^{20}$ atoms~cm$^{-2}$. As shown in the major-axis PV diagram (panel \textit{d)}), while $\mathrm{^{3D}}$Barolo models well the main rotating body, there is signal at around velocities of 10~km~s$^{-1}$ and offset 25~arcsec that cannot be reproduced by the best-fit model. The parameters for these galaxy are considered less robust than for the rest of the sample, as we discuss in Section~\ref{subsec:barolo}.}
    \label{fig:749290}
\end{figure*}{}

\subsection{Optical data}
Given the LSB nature of our galaxies, the imaging of wide-field public surveys like the Sloan Digital Sky Survey (SDSS, e.g., \citealt{york}) is not deep enough to provide accurate photometric parameters on an individual basis. Because of this, the six galaxies were observed using the One Degree Imager \citep{harbeck} of the 3.5-m WIYN telescope at the Kitt Peak National Observatory. The $g$ and $r$ bands were used, with a total exposure time of 45 min per filter. The optical image production is described in detail in \citet{gault}. Panel \textit{a)} in Figures~\ref{fig:114905}--\ref{fig:749290} shows the $r-$band optical images of our sample.

As introduced in \citet{btfr} and shown thoroughly in \citet{gault}, aperture photometry is performed on these images to obtain total magnitudes (and colours) and surface brightness profiles. The central surface brightness and disc scale length ($R_{\rm d}$) are obtained from a fit to the observed surface brightness profiles assuming that the light distribution follows an exponential profile, which is a good assumption for these galaxies (see also \citealt{roman2,greco,paperII}). 

\begin{table*}
\caption{Properties of our galaxy sample.}
\label{tab:properties}
\hspace{-8.5ex}
\begin{tabular}{ccccccccccccc}
	\hline
 ID & RA (J2000) & DEC (J2000) & $V_{\rm sys}$ & $D$ & $R_{\rm d}$ & $\mathrm{log(}M\mathrm{_\star / }\mathrm{M_\odot})$ & $\mathrm{log(}M\mathrm{_{HI} / }\mathrm{M_\odot})$ & Inc. & PA & $V_\mathrm{circ}$ & $\langle \upsigma \rangle$ & $R\mathrm{_{out}}$\\
  AGC & [hh:mm:ss.ss] & [dd:mm:ss.ss] & [km~s$^{-1}$] & [Mpc] & [kpc] &  &  & [deg] & [deg] & [km~s$^{-1}$] & [km~s$^{-1}$] & [kpc] \\
  (1) & (2) & (3) & (4) & (5) & (6) & (7) & (8) & (9) & (10) & (11) & (12) & (13) \\

   \hline
 114905 & 01:25:18.60 & +07:21:41.11 & 5435 &  76 & 1.79 $\pm$ 0.04 & 8.30 $\pm$ 0.17 &  9.03	$\pm$ 0.08  & 33 &  85 & 19$^{+6}_{-4}$ & $\lesssim$ 4 & 8.02 \\ \noalign{\smallskip}
 122966 & 02:09:29.49 & +31:51:12.77 & 6509 &  90 & 4.15 $\pm$ 0.19 & 7.73 $\pm$ 0.12 &  9.07	$\pm$ 0.05  & 34 & 300 & 37$^{+6}_{-5}$ & 7 & 10.80 \\ \noalign{\smallskip}
 219533 & 11:39:57.16 & +16:43:14.00 & 6384 &  96 & 2.35 $\pm$ 0.20 & 8.04 $\pm$ 0.12 &  9.21	$\pm$ 0.18  & 42 & 115 & 37$^{+5}_{-6}$ & $\lesssim$ 4 & 9.78 \\ \noalign{\smallskip}
 248945 & 14:46:59.50 & +13:10:12.20 & 5703 &  84 & 2.08 $\pm$ 0.07 & 8.52 $\pm$ 0.17 &  8.78	$\pm$ 0.08  & 66 & 300 & 27$^{+3}_{-3}$ & $\lesssim$ 4 & 8.55 \\ \noalign{\smallskip}
 334315 & 23:20:11.73 & +22:24:08.03 & 5107 &  73 & 3.76 $\pm$ 0.14 & 7.93 $\pm$ 0.12 & 9.10 $\pm$ 0.10  & 45 & 185 & 25$^{+5}_{-5}$  & 7 & 8.49  \\ \noalign{\smallskip}
 749290 & 09:16:00.95 & +26:38:56.93 & 6516 &  97 & 2.38 $\pm$ 0.14 & 8.32 $\pm$ 0.13 &  8.98	$\pm$ 0.08  & 39 & 130 & 26$^{+6}_{-6}$ & $\lesssim$ 4 & 8.47  \\
    \hline
\end{tabular}
    \begin{tablenotes}
      \small
      \item (1) Arecibo General Catalogue ID. (2-3) Right ascension and declination. (4) Systemic velocity. (5) Distance, taken from \citetalias{leisman}, has an uncertainty of $\pm$~5~Mpc. (6) Optical disc scale length, obtained from an exponential fit to the \textit{r}--band surface brightness profile. (7) Stellar mass. (8) H\,{\sc i} mass. (9) Inclination, derived from the H\,{\sc i} data with an uncertainty of $\pm 5^\circ$. (10) Kinematic position angle, derived from the H\,{\sc i} data, with an uncertainty of $\pm 8^\circ$. (11) Circular speed. (12) Mean value of the gas velocity dispersion . (13) Radius of the outermost ring of the rotation curve.
    \end{tablenotes}
\end{table*}

To derive the stellar masses we employ the mass-to-light--colour relation given by \citet{herrmann}:
\begin{equation} \label{eq:m2l}
    \log(M_\star/L_{\rm g}) = 1.294(\pm 0.401)\times (g-r) - 0.601(\pm 0.090)~,
\end{equation}
which was specifically calibrated for dwarf irregular galaxies, whose optical morphology is similar to isolated UDGs. 
In practice, for each UDG we randomly sample Equation~\ref{eq:m2l} using Gaussian distributions on each parameter, to account for the uncertainties in both the relation itself and the photometry. The fiducial value of each parameter is chosen as the mean of its Gaussian distribution while its standard deviation gives the corresponding uncertainty. With this, we obtain a distribution for $\log(M_\star/L_{\rm g})$ that is then converted to stellar mass using the $g-$band absolute magnitude distribution. The values for the stellar mass, which we report in Table~\ref{tab:properties}, are the median values of the final distributions for each galaxy, and the uncertainties the difference between these medians and the corresponding 16$^{\rm th}$ and 84$^{\rm th}$ percentiles. 

\subsection{H\,{\sc i} data}
We obtained resolved H\,{\sc i}-line observations using the Karl G. Jansky Very Large Array (VLA) and the Westerbork Synthesis Radio Telescope (WSRT). All the galaxies have radio data from the VLA, while AGC~122966 and AGC~334315 have also WSRT data. Details of the data reduction are given in \citetalias{leisman} and \citet{gault}. In the case of the two galaxies with VLA and WSRT observations, we use the data with the best quality in terms of spatial resolution and signal-to-noise ($S/N$), which were the VLA data for AGC~334315 and the WSRT data for AGC~122966. In the rest of this paper we use the parameters derived from these data. For completeness, in Appendix~\ref{app:extra} we present the WSRT data for AGC~334315\footnote{Note that \citet{btfr} used the WSRT data for AGC~334315 while we will use its VLA data for the rest of this work. Yet, the differences are rather small, as explained in Appendix~\ref{app:extra}.} and the VLA data for AGC~122966, demonstrating the overall good agreement between the different data cubes.

We build total H\,{\sc i} maps of our sources using the software $\mathrm{^{3D}}$Barolo (\citealt{barolo}, see below for more details). These maps are first obtained using a mask that $\mathrm{^{3D}}$Barolo generates after smoothing the data cubes by a given factor and then selecting those pixels above a chosen threshold in units of the rms of the smoothed cube. Upon inspection of our data, we find sensible values for the smoothing factor and the cut threshold around 1.2 and 3.5, respectively. The fluxes of our galaxies are measured from the data cubes using the task \textsc{flux} from \texttt{GIPSY} \citep{gipsy1, gipsy2}. 
The measurements of the flux from the VLA and WSRT data cubes are fully consistent with the separate analysis by \citet{gault} and \citetalias{leisman}, and in good agreement (within $\approx$10\%) with the values obtained from the ALFALFA single-dish observations, except for AGC~248945, for which we recover $\approx$30\% less flux. Upon inspection of the data cube, we confirm that the emission missing in the VLA profile with respect to ALFALFA is not biased with respect to the velocity extent of the source. Panel \textit{b)} in Figures~\ref{fig:114905}--\ref{fig:749290} presents the total H\,{\sc i} maps of our galaxies.

We determine the H\,{\sc i} mass of our UDGs using the equation
\begin{equation}
\frac{M\mathrm{_{HI}}}{\mathrm{M_\odot}} = 2.343 \times 10^5 ~ \bigg( \frac{D}{\textnormal{Mpc}} \bigg)^2 \bigg( \frac{F\mathrm{_{HI}}}{\textnormal{Jy km s$^{-1}$}} \bigg) ,
\end{equation}
with $D$ and $F\mathrm{_{HI}}$ the distance and flux of each galaxy, respectively. Distances, taken directly from \citetalias{leisman}, come from the ALFALFA catalog, which uses a Hubble flow model \citep{masters}. Given the line-of-sight velocities of our sample (see Table~\ref{tab:properties}), and considering that these UDGs live in the field, the possible effects of peculiar velocities are not significant and the Hubble flow distances provide a robust measurement of the `true' distance, with an uncertainty of $\pm$~5~Mpc.

\subsubsection{Interpretation of velocity gradients}

As can be seen in the panel \textit{c)} of Figures~\ref{fig:114905}--\ref{fig:749290}, we observe clear velocity gradients in most of our UDGs; AGC~749290 is the exception and the kinematics of this galaxy is more uncertain, as we discuss below. These gradients are along the morphological H\,{\sc i} position angle of the galaxies, and in the following sections we interpret them as produced by the differential rotation of a gaseous disc. Here we briefly discuss other possibilities.

One may wonder if the observed velocity gradients could be generated not by rotation but by gas inflow (see \citealt{renzo} for a review) or blown-out gas due to powerful stellar winds (see for instance \citealt{mcquinn} and references therein). Such winds have been observed in starburst dwarfs traced by H$\alpha$ emission, where the H\,{\sc i} distribution may also be disturbed (e.g., \citealt{lelli2014,mcquinn}). 
There is, however, clear evidence against these scenarios in the case of our UDGs. First of all, the velocity gradients are aligned with the H\,{\sc i} morphological position angle of the galaxies, as happens with normal rotating discs. Further, as we discuss later, our measurements of the gas velocity dispersions point to a rather undisturbed and quiet ISM. Moreover and most importantly, our galaxies have normal-to-low star formation rates (SFR $\approx$ 0.02--0.4~M$_\odot$~yr$^{-1}$, see \citetalias{leisman}), which combined with their extended optical scale length leads to SFR surface densities of a factor about $10-20$ lower than in typical dwarfs. The fact that our UDGs are gas dominated and there is only one clear velocity gradient implies that if the gradients are due to winds the whole ISM should be in the wind, requiring very high mass loading factors and SFR densities, in contradiction with the information presented above. Based on this discussion we conclude that the possibility of the observed velocity fields being produced by inflows or outflows is very unlikely. In contrast, we show in Section~\ref{sec:kinematics} how a rotating disc can reproduce the features observed in our data.

\subsection{Baryonic mass}
The baryonic masses of the galaxies are computed with the equation
\begin{equation}
M_{\rm bar} = M_{\rm gas} + M_\star = 1.33~M_{\rm HI} + M_\star~,
\label{eq:mass}
\end{equation}
where the factor 1.33 accounts for the presence of helium.

The mass budget of our galaxies is dominated by the gas content, with a mean gas-to-stellar mass ratio ($M\mathrm{_{gas}/}M\mathrm{_{\star}}) \approx$~15 (see \citealt{btfr} for more details). This ensures that, despite possible systematics when deriving the stellar mass, the estimation of the baryonic mass is robust. 

As seen in Eq.~\ref{eq:mass}, we neglect any contribution from molecular gas to the baryonic mass of the galaxies; while the molecular gas mass is indeed often smaller than the stellar and atomic gas ones in dwarfs (e.g. \citealt{leroy, saintonge, anastasia2}, and references therein), it may of course contribute to the total mass budget. This hypothetical baryonic mass gain, however, would place our sources further off the BTFR, only strengthening the results shown in Section~\ref{sec:btfr}.

\section{Deriving the gas kinematics}
\label{sec:kinematics}
\subsection{Initial parameters for 3D modelling}
Our interferometric observations allow us to estimate rotation velocities for the six galaxies. However, the data have low spatial resolution, with only a couple of resolution elements per galaxy side. Low-spatial resolution observations can be severely affected by beam smearing, which tends to blur the observed velocity fields, and traditional 2D approaches that fit tilted-ring models to beam-smeared velocity fields fail at recovering the correct kinematics, by underestimating the rotation and overestimating the gas velocity dispersion (e.g., \citealt{bosma,swaters,enrico2}).

$\mathrm{^{3D}}$Barolo\footnote{Version 1.4, \url{http://editeodoro.github.io/Bbarolo/}.} \citep{barolo} is a software tool which produces 3D models of emission-line observations (e.g., \citealt{enrico2,iorio, ceci}). Instead of fitting the velocity field, it builds 3D tilted-ring realizations of the galaxy that are later compared with the data to find the best-fit model. Thanks to a convolution step before the model is compared with the data, $\mathrm{^{3D}}$Barolo strongly mitigates the effect of beam smearing, so it is ideal for analysing data like ours. $\mathrm{^{3D}}$Barolo assumes that the discs are thin; while this is not known a priori, we show in Section~\ref{sec:thickness} that the ratio between the radial and vertical extent of our UDGs is large, confirming the validity of our approach. 

Due to the small number of resolution elements we prefer to fit only two parameters with $\mathrm{^{3D}}$Barolo: the rotation velocity and the velocity dispersion. This means that the rest of the parameters need to be determined and fixed, namely the center of the galaxy, its systemic velocity, position angle and inclination.

$\mathrm{^{3D}}$Barolo can robustly estimate the systemic velocity and the centre of the galaxies from the centre of the global H\,{\sc i} profile and the total H\,{\sc i} map, respectively. We use thus these estimations from $\mathrm{^{3D}}$Barolo and we keep them fixed while fitting the rings. $\mathrm{^{3D}}$Barolo can also estimate the position angle and inclination, but for low resolution data these estimates may not be accurate. Therefore, we decide to estimate these two parameters independently and to fix them when fitting the kinematic parameters.
The position angle is chosen as the orientation that maximizes the amplitude of the position-velocity (PV) diagram along the major axis. This is done visually using the task \textsc{kpvslice} of the \textsc{karma} package \citep{karma}. Importantly, we find that in every galaxy the kinematic and morphological (H\,{\sc i}) position angle are nearly the same. The exception may seem to be AGC~122966, but as we discuss in Section~\ref{subsec:barolo} this is an apparent artifact due to the shape of the beam for the WSRT observations of that galaxy. 

Estimating the inclination of the galaxies is of crucial importance, as correcting for it can account for a large fraction of the final rotation velocity if the galaxies are seen at low inclinations.
Unfortunately, due to the LSB nature of our galaxies, their optical morphologies, often irregular and dominated by patchy regions, provide only an uncertain, if any, constraint on the inclinations (see also \citealt{starkenburg} for other limitation of using optical data to determine the inclination of the H\,{\sc i} disc.). This, together with the fact that the H\,{\sc i} is more extended and massive than the stellar component, motivated us to use the H\,{\sc i} maps to estimate the inclinations. We do this by minimizing the residuals between the observed H\,{\sc i} map of each galaxy and the H\,{\sc i} map of models of the same galaxy but projected at different inclinations between 10$^\circ$ and 80$^\circ$. Such models are produced using the task \textsc{GALMOD} from $\mathrm{^{3D}}$Barolo, which in turn uses updated routines from the homonym \texttt{GIPSY} task, and takes into account the shape of the beam when generating the models. The centre, surface density and position angle for the models are the same as in the galaxy whose inclination we aim to determine. The inclination of the model that produces the lowest residual when compared with the data (lowest absolute difference between the total H\,{\sc i} maps) is chosen as the fiducial inclination.

\subsubsection{Testing on simulated galaxies}
\begin{figure}
    \centering
    \includegraphics[scale=0.422]{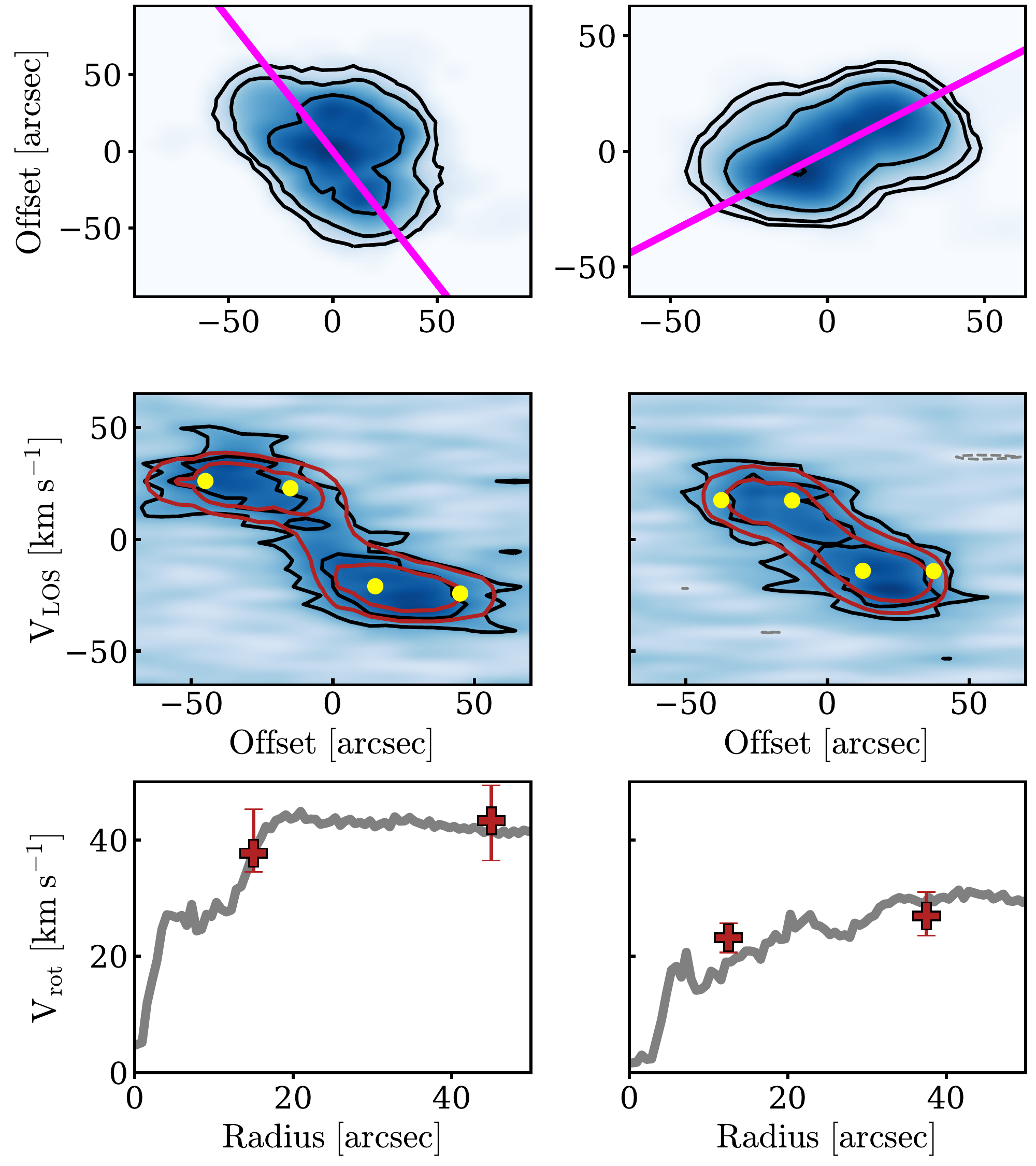}
    \caption{Two of the 40 \texttt{APOSTLE} simulated dwarf galaxies used to test our methods. \textit{Top}: Total H\,{\sc i} maps with their kinematic major axes in magenta. \textit{Middle}: PV diagrams along the major axis. The black and red contours represent the simulated data and our best-fit model, respectively; grey dashed contours show negative values. The yellow points show the recovered rotation velocities. \textit{Bottom}: True rotation curves (grey lines) derived directly from the simulations and recovered rotation velocities (red crosses).}
    \label{fig:simulated}
\end{figure}

We test our method to recover the galaxy initial geometrical parameters using a sample of gas-rich dwarfs from the \texttt{APOSTLE} cosmological hydrodynamical simulations \citep{apostle2, apostle1}. Mock H\,{\sc i} data cubes of these galaxies, `observed' at a resolution and $S/N$ matching our data, are obtained with the software {\sc martini}\footnote{Version 1.0.2, \url{http://github.com/kyleaoman/martini}.} \citep{kyle2019}. The simulated galaxies have H\,{\sc i} masses of $\sim$10$^{8-9}$~M$_\odot$ and rotation velocities around $\approx 20-60$~km~s$^{-1}$. We initially work with four simulated galaxies with similar mass and velocity as our sample, but we project them at different random position angles and inclinations, allowing us in practice to test our methods in 40 different mock data cubes. Figure~\ref{fig:simulated} shows two examples of such simulated galaxies: their H\,{\sc i} maps, PV diagrams, and rotation curves.

\begin{figure}
    \centering
    \includegraphics[scale=0.415]{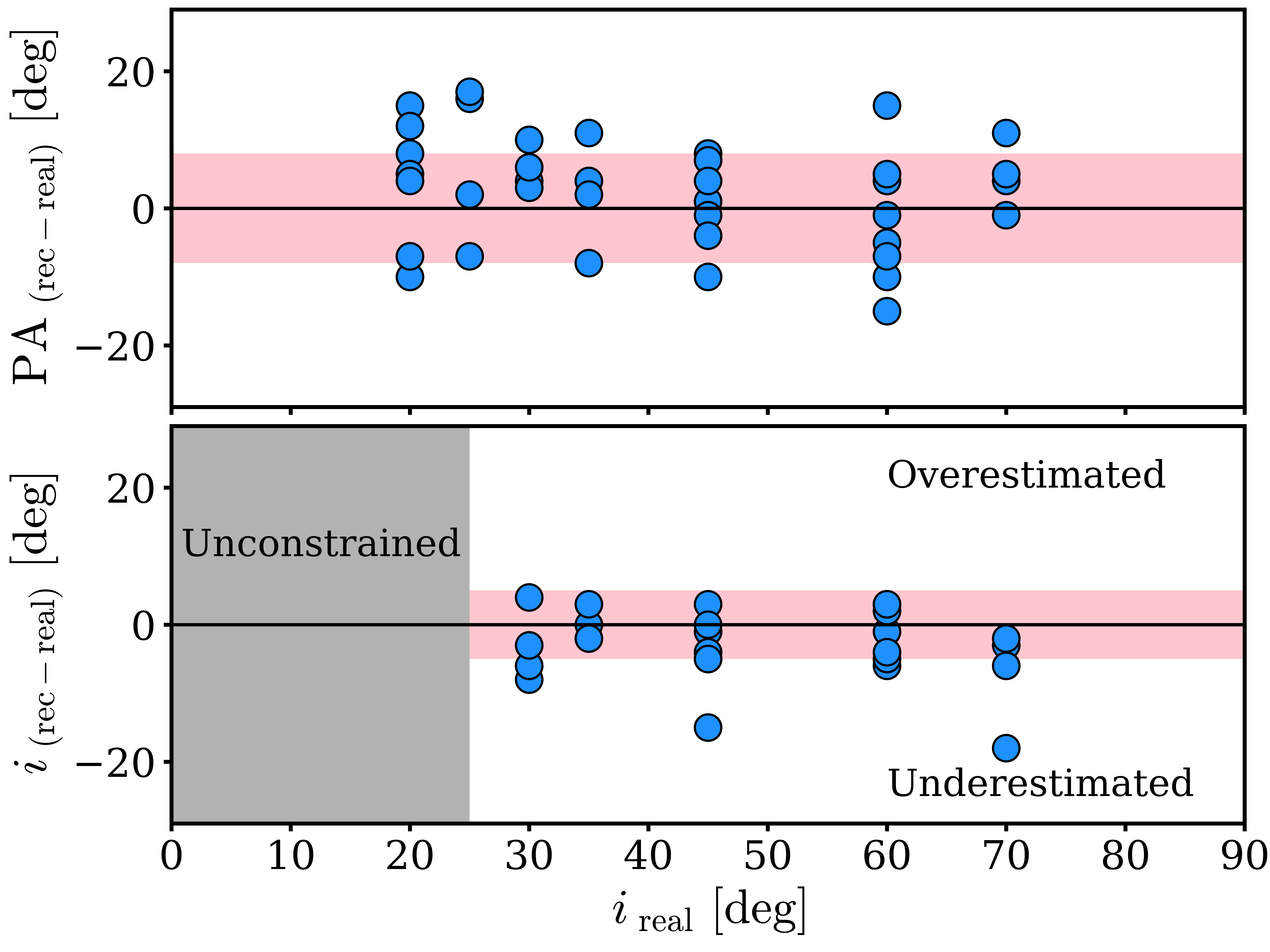}
    \caption{Comparison between truth (real) and recovered (rec) position angles (PA) and inclinations ($i$) in our set of \texttt{APOSTLE} dwarfs. The black lines show the case where the difference is zero, and the pink bands show our adopted uncertainties in both parameters.}
    \label{fig:inclinations}
\end{figure}

We treat these mock data in exactly the same way as our UDGs data, using the method described above to derive the position angle and inclination. Figure~\ref{fig:inclinations} shows the true and recovered geometrical parameters. We find that we can consistently recover the position angle of the simulated galaxies and, once this is fixed, the inclination. The mean of the absolute difference between truth and recovered position angles and inclination angles is 8$^{\circ}$ and 5$^{\circ}$, respectively. We adopt these values as the uncertainties for these parameters. Note that our method recovers the inclination only for galaxies with inclinations $\gtrsim 25^\circ$; below this all the models look very similar and our method systematically underestimates the inclination by about 10$^\circ$. For higher inclinations, in two out of 40 cases we underestimate the inclination by about 15$^\circ$. Given the low incidence of this (5 per cent of the times) we do not expect to underestimate the inclination of our real galaxy sample; in any case this would lower the circular velocities of our sample that we report, which would not affect the nature of our results below. The bottom panels of Figure~\ref{fig:simulated} show that our derived rotation velocities well represent the underlying rotation curves after having estimated the position and inclination angles as described above, and then used $\mathrm{^{3D}}$Barolo.

\subsection{Running $\mathrm{^{3D}}$Barolo on the individual systems}
\label{subsec:barolo}
After testing our methods we proceed to run $\mathrm{^{3D}}$Barolo on all our UDGs. As discussed before, we leave the rotation and dispersion as free parameters, and we fix the position angle and inclination; the values of these parameters are given in Table~\ref{tab:properties}. As expected, $\mathrm{^{3D}}$Barolo is able to estimate the centre and systemic velocity of the sources with good accuracy, as we could verify by visually inspecting the velocity fields and PV diagrams. 
It is worth mentioning that the systemic velocities in the kinematic fits agree well with the values determined from the ALFALFA global profiles (the mean difference is about 5~km~s$^{-1}$). The kinematic centres and systemic velocities used in the models can be found in Table~\ref{tab:properties}. The noise in the data cubes and peak column densities are provided in \citet{gault}. $\mathrm{^{3D}}$Barolo also applies a correction for instrumental broadening, controlled by the parameter \texttt{LINEAR}, which depends on the spectral smoothing of the data. In our case, we use \texttt{LINEAR}~=~0.42 for the VLA data, and \texttt{LINEAR}~=~0.85 for the (Hanning-smoothed) WSRT data. 

The separation between rings is given by the parameter \texttt{RADSEP} in $\mathrm{^{3D}}$Barolo, and the value is chosen taking into account the beam orientation and extension for each galaxy. Below we provide information on the value of \texttt{RADSEP} used for each galaxy, as well as some individual comments.

\begin{itemize}
    \item AGC 114905: The size of the beam is 14.64"$\times$13.31", with a North-West orientation of $-5^\circ$. The galaxy position angle is 85$^\circ$. The component of the beam projected along the kinematic major axis has a size of approximately the size of the beam minor axis. Given the extension of the galaxy, we use two rings, with \texttt{RADSEP} = 14.5".

    \item AGC 122966: The size of the beam is 33.16"$\times$18.70", oriented at 15$^\circ$. Given the orientation of the galaxy, the component of the beam along its kinematic major axis is $\approx$~20.5". Considering the extension of the galaxy ($\approx$ 33") we oversample by a factor $\approx$~1.2, using \texttt{RADSEP} = 16.5" and allowing us to have two points.
    This galaxy gives the impression of having perpendicular morphological and kinematic major axes, but this is an apparent effect due to the elongated shape of the WSRT beam, as it can be seen in Appendix~\ref{app:extra} with the less elongated beam of the VLA data.

    \item AGC 219533: Beam size of 14.93"$\times$13.62", with orientation at 25.5$^\circ$. The projected beam radius along the kinematic major axis of the galaxy is 13.6". We use two independent resolution elements with \texttt{RADSEP} = 14".

    \item AGC 248945: Beam size of 18.10"$\times$14.11", with a position angle of 57.3$^\circ$. The projected beam radius along the kinematic major axis of the galaxy is 14.7". Given the extension of the H\,{\sc i} emission we adopt \texttt{RADSEP} = 14.5", ending up with two resolution elements.
    
    \item AGC 334315: The galaxy has a beam size of 15.83"$\times$13.94", oriented at $-65^\circ$. Along the major axis of the galaxy, the projected size of the beam is $\approx$ 14". We use two independent resolution elements with \texttt{RADSEP} = 16". 
    
    \item AGC 749290: Beam size of 21.63"$\times$17.88", oriented at $-61^\circ$. The projected radius of the beam along the major axis of the galaxy is $\approx$~21.4". Given the extension of the galaxy we oversample by a factor 1.7 to get two resolution elements per galaxy side, with \texttt{RADSEP} = 12", although the resulting two rings are not independent. Because of this, the kinematic parameters of this galaxy are less certain than for the rest of our sample, and we plot the galaxy as an empty symbol when using its kinematic parameters. Nevertheless, we note that the specific values of its circular speed and velocity dispersion are similar to the values of the rest of the sample.
\end{itemize}

\subsection{Kinematic models}

For all our galaxies the kinematic fits converge and $\mathrm{^{3D}}$Barolo finds models which are in good agreement with the data. Figures~\ref{fig:114905}--\ref{fig:749290} show our kinematic models in panels \textit{c}) to \textit{f)}: observed and modelled velocity field, and observed and modelled (1 pixel width) PV diagrams.

The PV diagrams and rotation velocities suggest that we are tracing the flat part of the rotation curve, as the two points of the rotation curves are consistent with each other. This may be a possible source of confusion since some PV diagrams, at first-sight, may look like solid-body rotation. However, this is an effect of the beam smearing, and $\mathrm{^{3D}}$Barolo is able to recover the intrinsic rotation velocities (see for instance figure~7 and 8 in \citealt{barolo}), although for AGC~749292 this is not possible to establish unambiguously as we oversample the data by a factor 1.7. Moreover, standard rotation curves of simulated dwarf galaxies are expected to reach the flat part well inside our typical values of $R_{\rm out}$ (e.g., \citealt{kyle2015}). Observed rotation curves do not keep rising after 2-3$R_{\rm d}$ either (e.g., \citealt{swaters}), which is again inside our values of $R_{\rm out}$. Yet, higher-resolution and higher-sensitivity observations would be desirable to further confirm this, as well as to trace the inner rising part which we cannot observe at the current resolution.

Taking this into account, and the fact that the inclination is the main driver of uncertainties in the rotation velocity, we estimate the circular speeds and their uncertainties reported in Table~\ref{tab:properties} as follows:
\begin{enumerate}
    \item For each galaxy, we run $\mathrm{^{3D}}$Barolo two more times, but instead of using our fiducial inclination $i$, we use $i-5$ and $i+5$. This means that each ring of a galaxy has three associated velocities, obtained with $i$, $i-5$ and $i+5$. $\mathrm{^{3D}}$Barolo is able to correct for pressure-supported motions, with the so-called asymmetric drift correction (e.g., \citealt{iorio}), allowing the conversion from rotation velocities to circular speeds. We apply this correction, although it is found to be small, contributing at most $\approx 1$~km~s$^{-1}$. 
    
    \item For each of the above velocities (at $i$, $i-5$ and $i+5$), and for each ring, we generate random Gaussian distributions centred at the value of the velocity, and with standard deviation given by the statistical errors in the fit found by $\mathrm{^{3D}}$Barolo. A galaxy with two resolution elements has six corresponding Gaussian distributions, three for each ring.
    
    \item Finally, we add all these Gaussian distributions in a broader distribution $\mathcal{G}$. For each galaxy, the circular speed ($V_{\rm circ}$) corresponds to the 50$^{\rm {th}}$ percentile of $\mathcal{G}$, and its lower and upper uncertainties (Table~\ref{tab:properties}) correspond to the difference between that value and the 16$^{\rm {th}}$ and 84$^{\rm {th}}$ percentiles of $\mathcal{G}$, respectively.
\end{enumerate}

\subsubsection{Circular speeds}
\label{sec:circularspeeds}
Our galaxies have circular speeds between 20 and 40~km~s$^{-1}$. Given their baryonic masses, their velocities are a factor $2-4$ lower than the expectations from the BTFR (see Section~\ref{sec:btfr} and \citealt{btfr}). Our lower-than-average circular speeds are consistent with earlier observations by different authors that these kinds of galaxies have narrower global H\,{\sc i} profiles than other galaxies with similar masses (\citetalias{leisman}; \citealt{spekkens, jano}).

A question that may arise, given the long dynamical timescales implied by the low rotation velocities of our UDGs, is whether they are in dynamical equilibrium. The average dynamical time for our sample is 2~Gyr. The mean distance from our UDGs to their nearest neighbor, according to the Arecibo General Catalog\footnote{The Arecibo General Catalog is a catalogue containing all the sources detected in the ALFALFA survey plus all the galaxies with optical spectroscopic detections within the ALFALFA footprint. It is compiled and maintained by Martha Haynes and Riccardo Giovanelli.}, is 1~Mpc. If we consider the case where \textit{all} our galaxies interacted with their nearest neighbor, and we assume that they come from a 10$^{12}$~M$_\odot$ environment (gas-rich UDGs inhabit low-density large-scale environments, see \citealt{jano}) with an escape speed of 200~km~s$^{-1}$, the mean interaction back-time (how long ago did the interaction occur) for them is about 5~Gyr, so the galaxies should have had time to reach a stable configuration, having completed on average more than two full rotations. 

\subsubsection{Velocity dispersion}
\label{sec:dispersion}
The narrowness of the (beam-smeared) PV diagrams of our galaxies, shown in Figures~\ref{fig:114905}--\ref{fig:749290}, suggests a rather low gas velocity dispersion for most of them. This is indeed confirmed by the best-fit models of $\mathrm{^{3D}}$Barolo. The mean velocity dispersion for AGC~114905, AGC~219533, AGC~248945 and AGC~749290, with a channel width of $\Delta_{\rm v} \approx$~4~km~s$^{-1}$, is $\langle \upsigma \rangle = 3 \pm 2 $~km~s$^{-1}$, which is below $\Delta_{\rm v}$. However, based on tests using artificial data cubes, we find that, for data like ours, $\mathrm{^{3D}}$Barolo cannot recover the exact velocity dispersion if this lies below $\Delta_{\rm v}$, but it tends to find $\langle \upsigma \rangle \approx \Delta_{\rm v}$. Therefore, for these data cubes we assume an upper limit of $\langle \upsigma \rangle \lesssim$~4~km~s$^{-1}$. For AGC~334315, with the same $\Delta_{\rm v} \approx$~4~km~s$^{-1}$, we find $\langle \upsigma \rangle = 7 \pm 2 $~km~s$^{-1}$, and we adopt this value. The WSRT data for AGC~122966, which is Hanning smoothed and of lower spatial resolution ($\Delta_{\rm v} \approx$~6~km~s$^{-1}$), has $\langle \upsigma \rangle = 7 \pm 2$~km~s$^{-1}$.

The observed gas velocity dispersions are lower than observed in typical spiral and dwarf galaxies. The upper limit in the velocity dispersion of the VLA cubes is indeed similar to the velocity dispersion of the `cold' neutral medium of Leo~T \citep{leot}. For comparison, \citet{iorio} in their reanalysis of the kinematics of dwarf galaxies from LITTLE THINGS \citep{hunterLT} found $\langle \upsigma \rangle \sim$~9~km~s$^{-1}$, similar to the $\langle \upsigma \rangle \sim$~10~km~s$^{-1}$ of both the more massive spirals of \citet{tamburro} and \citet{ceci} and the regularly-rotating starburst dwarfs from \citet{lelli2014}. In the next section we explore the repercussions of these results. 

\section{Thickness and turbulence in the discs of gas-rich UDGs}
\label{sec:ISM}
\subsection{Thickness of the gas disc}
\label{sec:thickness}
Given the gravitational potential and gas surface density of a galaxy, the value of its velocity dispersion can be used to estimate its gas disc scale height $h$ (see for instance \S~4.6.2 in \citealt{bookFilippo}). Since our galaxies are dominated by the gas component rather than the stellar and dark matter components, at least up to the outermost measured point (see figure~3 in \citealt{btfr}), we can consider the simple case of a self-gravitating disc with constant circular speed in hydrostatic equilibrium (e.g., \citealt{vanderkruit,marascoFF}). This exercise only provides an indicative value for $h$, but it is still instructive as this measurement has not been yet carried out for UDGs. The scale height of such discs is given by the equation
\begin{equation}
    h =  \dfrac{\upsigma^2}{\pi G \Sigma_{\rm{gas}}}~,
\end{equation}
with $\upsigma$ the gas velocity dispersion, $G$ the gravitational constant and $\Sigma_{\rm{gas}}$ the gas surface density. 

Assuming a mean velocity dispersion constant with radius and the mean surface density of the disc\footnote{We do this for simplicity, ending-up with a constant scale height, but our discs may be flared as in other dwarfs and spiral galaxies (e.g., \citealt{banerjee, ceci}).}, we obtain a mean (median) disc scale height of $\langle h \rangle = $~260 (150)~pc. Note that these values may in reality be smaller, as \textit{i)} we are adopting an upper limit in the velocity dispersion for most galaxies, and \textit{ii)} we completely neglect the potential provided by the stars and dark matter, which, even if small, would contribute to flatten the disc. We can conclude that our galaxies do not appear to have H\,{\sc i} discs significantly thicker than other disc galaxies. For reference, the H\,{\sc i} discs studied in \citet{ceci} have mean values for $h$ between $130-540$~pc, depending on the galaxy, and the dwarfs from \citet{banerjee} have $\langle h \rangle \approx 500$~pc. 
Note that the differences in the assumed shape of the vertical profile are not very big: for instance, the correction for using sech$^2$ instead of a Gaussian function (as in \citealt{ceci}) is less than 10\% of the value of $h$.

It is also worth highlighting that given the H\,{\sc i} radius ($R_{\rm HI}$) of our sample \citep{gault}, the values of $h$ indicate that they have relatively `thin' discs: the extension of their discs (using for reference $R_{\rm HI}$) is on average 50 times larger than the size of the scale height. This result also confirms that our approach of using $\mathrm{^{3D}}$Barolo (where galaxies are modelled as thin discs) is perfectly adequate.

\subsection{Turbulence in the ISM}
According to the \citet{field} criterion for thermal instabilities, the ISM should only exist in stable conditions in two well-defined phases. These two phases correspond to the cold (CNM) and warm neutral media (WNM), with temperatures of $\sim 70-100$~K and $\sim 6000-8000$~K, respectively, although in realistic conditions gas in the interfaces of both media exists at intermediate temperatures (e.g., \citealt{heiles}). These temperatures imply a thermal speed of $0.75-1$~km~s$^{-1}$ for the CNM and $7-8$~km~s$^{-1}$ for the WNM. 
 
In this context, our UDGs are an intriguing case because the observed intrinsic velocity dispersions are lower than the expected thermal speed of the WNM. Assuming that indeed the galaxies lack of a significant amount of WNM, the velocity dispersion can be then attributed entirely to the thermal broadening of the CNM plus turbulence in the disc\footnote{Note that neutral gas at $T \sim 2000$~K can produce a dispersion of 4~km~s$^{-1}$, without additional energy input.}. By further assuming that turbulence is driven entirely by supernova explosions, we can compute the supernova efficiency in transferring kinetic energy to the ISM (see for instance \S~8.7.4 in \citealt{bookFilippo} or \S~VI in \citealt{maclow}). We find that efficiencies between 2 and 5 per cent are enough to reproduce the observed low gas velocity dispersions. While these values are limited by all our uncertainties and are valid only within about one order of magnitude, they indicate that the supernova efficiency in our UDGs is likely similar to the expectations for disc galaxies from different theoretical papers like \citet{thornton,fierlinger} or recent observational results (Bacchini et al. submitted to A\&A), but different from the results reported in other observational works like \citet{tamburro} or \citet{utomo}, where supernova efficiency needs to be very high and even external drivers of turbulence (e.g., magneto-rotational instabilities) are needed. Overall our discussion here and in Sec.~\ref{sec:dispersion} highlights the `cold' nature of the H\,{\sc i} disc of our UDGs. 

\section{Understanding the deviation from the BTFR}

\label{sec:btfr}
As discussed in \citet{btfr}, our UDG sample lies off the canonical BTFR, with circular speeds $2-4$ times lower than galaxies with similar masses or, equivalently, with $10-100$ times more baryonic mass than galaxies with similar circular speeds. This result holds after taking into account all the different possible systematics while deriving the circular speeds and baryonic masses. 

\begin{figure*}
    \centering
    \includegraphics[scale=0.65]{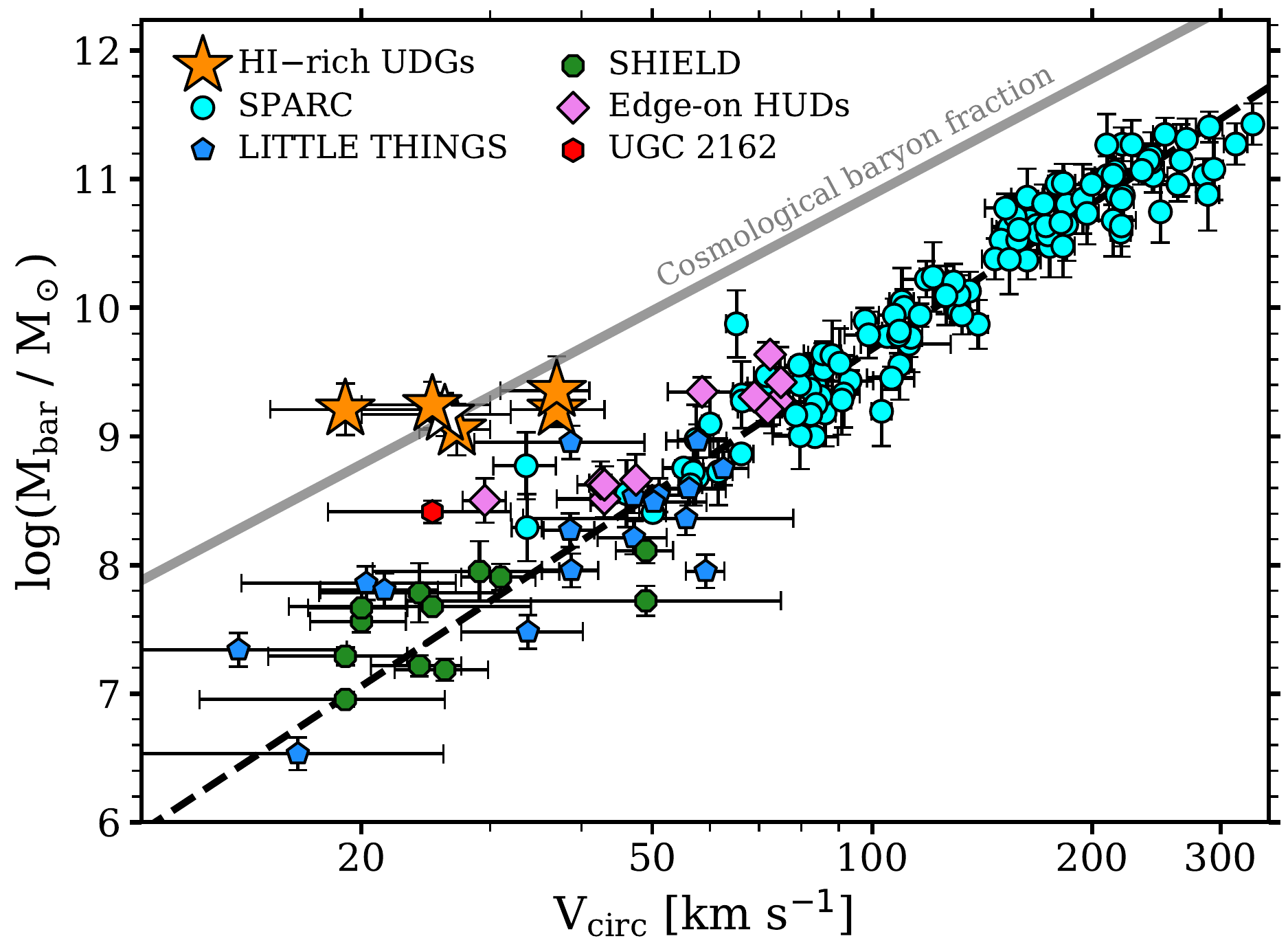}
    \caption{Circular speed vs. baryonic mass plane for different galaxy samples. Gas-rich UDGs are shown as orange stars, except for AGC~749290, whose circular speed is less robust than for the rest of the sample, and it is shown as a white star. The black dotted like shows the fit to the SPARC galaxies from \citet{lelli2016}, extrapolated towards low circular speeds. The grey solid line is the expectation for galaxies that have a baryon fraction equal to the cosmological mean. When including more galaxies from the literature in this plane, the apparent gap between the canonical BTFR and our UDGs is populated. See the text for details.}
    \label{fig:BTFRall}
\end{figure*}

\begin{figure}
    \centering
    \includegraphics[scale=0.55]{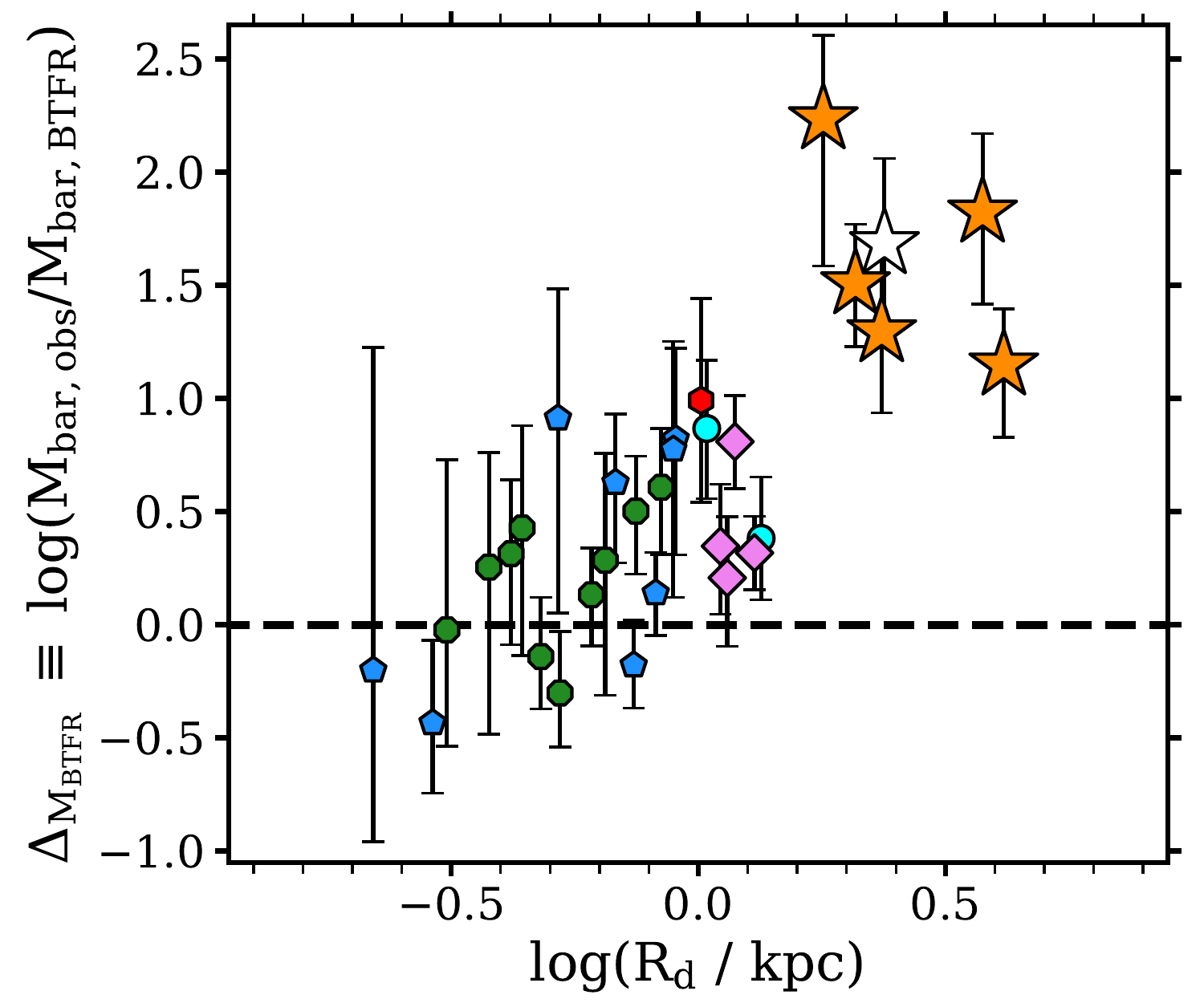}
    \caption{Disc scale length vs. vertical distance from the BTFR, for galaxies of different samples with 15~km~s$^{-1}< V_{\rm circ} < 45$~km~s$^{-1}$. Symbols are as in Figure~\ref{fig:BTFRall} and the dashed line represents no offset from the SPARC BTFR. A correlation between both parameters is observed, with larger galaxies falling systematically above the BTFR. Some samples have no reported uncertainty in $R_{\rm d}$, so we do not plot any horizontal error-bar for consistency.}
    \label{fig:VD_BTFR}
\end{figure}

\citet{btfr} postulate that it may not be surprising that no other galaxies have been found to lie on a similar position off the BTFR as interferometric observations are usually targeted based on optical detections and the UDGs are a faint optical population. Some galaxies in the literature (e.g., \citealt{geha,kirby12,kyle2016}) also appear to be outliers from the BTFR, although concerns regarding their kinematic parameters have been raised (see discussion in \citealt{kyle2016}). In this section we study in more detail the existence of outliers from the BTFR, using more galaxies with resolved 21-cm observations from the literature than in \citet{btfr}. We plot our UDGs (stars) and the different comparison samples in Figure~\ref{fig:BTFRall}, together with the best-fit line to the SPARC galaxies from \cite{lelli2016}, extrapolated towards the low-circular speed regime, and the expected relation for galaxies with a baryon fraction equal to the cosmological mean (see \citealt{btfr}). The reader interested in the main results of this comparison, without delving into the details, may wish to go ahead to Figures~\ref{fig:BTFRall}, \ref{fig:VD_BTFR} and Section~\ref{sec:btfr}. 

We start by considering all the galaxies from the SPARC sample \citep{sparc}. From the 175 galaxies listed in the data base\footnote{\url{http://astroweb.cwru.edu/SPARC/SPARC_Lelli2016c.mrt}}, 135 have an available measurement of their asymptotically flat rotation velocity. Five out of these 135 galaxies are included in the LITTLE THINGS sample of \citet{iorio}, and since their analysis is more detailed and similar to ours we do not use the SPARC values for these galaxies. From the remaining 130 galaxies we select those with inclinations $i \geq$~30$^\circ$ and good quality flag on their rotation curve ($Q$ = 1, 2, see \citealt{sparc} for details), ending up with 120 galaxies, shown in Figure~\ref{fig:BTFRall} as cyan circles.

We consider also the LITTLE THINGS galaxies from \citet{iorio}, shown in Figure~\ref{fig:BTFRall} as blue pentagons, and the SHIELD galaxies \citep{shield}, plotted as green octagons. Additionally, we include UGC~2162 (red hexagon), a UDG with resolved GMRT data presented in \citet{sengupta}, and a sample of nearly edge-on `H\,{\sc i}--bearing ultra-diffuse sources' (HUDs, see \citetalias{leisman}) with ALFALFA data from \citet{he}, shown as magenta diamonds. 

While we restrict our comparison to samples with resolved H\,{\sc i} data and the sample of UDGs from \citet{he}, some other studies based on unresolved H\,{\sc i} data are also worth briefly mentioning in the context of the BTFR. For example, the sample from \citet{geha} shows a number of low-mass dwarf galaxies that increase the scatter of the BTFR at low velocities. In particular, for $V_{\rm rot} \lesssim$~40~km~s$^{-1}$, most of their dwarfs rotate too slowly for their baryonic mass (see their figure~7). 
Also, \citet{guo} have recently used unresolved data from ALFALFA and faint SDSS imaging to suggest that 19 dwarfs they observe show similar properties as those discussed in \citet{btfr}: the galaxies seem to rotate too slowly for their masses and to have less dark matter than expected. The result from \citet{guo} by itself, however, may be subject of concern as they used unresolved data to estimate the rotation velocities, and lack information on the inclination of the H\,{\sc i} disc as they derive an inclination from shallow SDSS data that may not inform us on the actual orientation of the disc (see for instance \citealt{starkenburg, sanchezalmeida} and \citealt{gault}).\\

\noindent
The outcome of including all the different samples can be seen in Figure~\ref{fig:BTFRall}. Appendix~\ref{app:comparisonsamples} provides comments on the most interesting individual galaxies from each of the samples discussed above. 
In general, Figure~\ref{fig:BTFRall} suggests that it is likely that our UDGs are not the only outliers from the canonical BTFR at low circular speeds, although they may be the most extreme cases. In this context, we examine the deviation from the SPARC fit as a function of central surface brightness and disc scale length; in Appendix~\ref{app:comparisonsamples} we provide the references from which the structural parameters of the galaxies in Figure~\ref{fig:BTFRall} are obtained. 

We realize that those galaxies above the SPARC fit usually have lower surface brightness than galaxies in the relation, as expected for a constant $M/L$ (see discussion in \citealt{zwaan95} and \citealt{mcgaugh98}). However, this is not true for all the galaxies, and the analysis may be significantly influenced by the different strategies employed to derive the surface brightness in the literature (e.g., if values are corrected or not for inclination, dust reddening and Galactic extinction, and if different filters were used). Instead, measuring the radius of galaxies is more straightforward, as it has been shown to be less dependent on the different optical and infrared bands used to derive it (see for instance figure~B2 in \citealt{roman} and figure~1 in \citealt{falconbarroso}).

Different authors have found no correlation of the residuals of the best-fit BTFR and observations with other galaxy parameters. For instance, \citet{lelli2016} reported no trend as a function of effective radius, scale length or central surface brightness, and \citet{anastasia2} extended these results for Hubble type, colour, SFR and gas fraction (see also \citealt{anastasia} and references therein). Notwithstanding, \citet{vladimir}, with a larger fraction of LSB galaxies, reported that the scale lengths of their sample do correlate with the residuals of the BTFR, with smaller galaxies deviating towards higher velocities at fixed baryonic mass (note however that they looked at the BTFR using $V_{\rm max}$, instead of $V_{\rm flat}$ like in the other two mentioned studies). Apart from the existence of these discrepancies, those works include only a few galaxies with circular speeds similar to those of our UDGs ($V_{\rm circ} \approx 20-40$~km~s$^{-1}$), so it is interesting to re-consider the possible existence of correlations within the same range in velocity as our sample, using the compilation of galaxies that we have shown in this section.

Given the values of $V_{\rm circ}$ for our sample, we consider galaxies with 15~km~s$^{-1}< V_{\rm circ} < 45$~km~s$^{-1}$, and we use them in Figure~\ref{fig:VD_BTFR} to build the $R_{\rm d}-\Delta_{\rm M_{\rm bar}}$ plane. Here $R_{\rm d}$ is the stellar scale length and $\Delta_{\rm M_{\rm bar}}$ the vertical distance of the galaxies from the BTFR, defined as the logarithmic difference between the observed baryonic mass and the value expected from the extrapolated SPARC BTFR, $\Delta_{\rm M_{\rm bar}} \equiv \log(M_{\rm {bar,obs}}/M_{\rm{bar,BTFR}})$. A clear trend is found: at these low circular speeds, larger (more diffuse) galaxies lie systematically above the SPARC BTFR while the more compact ones lie below; the correlation has a slope around 1.5.

Spearman tests tell us that the correlation is significant at a 99.9 per cent confidence level ($p-$value $\approx 10^{-8}$) when all the samples are considered. This holds even if we exclude our UDGs ($p-$value $\approx 0.0003$). The correlation is less robust but still significant at the 95 per cent level when considering exclusively the SPARC and LITTLE THINGS galaxies ($p-$value $\approx 0.02$).

This trend is potentially of great importance because it provides evidence supporting the idea that the deviation from the BTFR at low circular speeds is driven by physical processes related to the optical size of the galaxies (which is independent of the kinematics), and that it is not only an effect produced by observational biases. 

One may wonder whether it is possible to interpret the trend as a spurious relation due to a severe underestimation of the circular speed of the galaxies: if the galaxies that deviate from the SPARC BTFR have wrong measurements but actually have $V_{\rm circ} \approx 80-100$~km~s$^{-1}$, then they would be expected to have larger scale lengths, giving rise to the trend observed in Figure~\ref{fig:VD_BTFR}. 

We find this unlikely for several reasons. First, as discussed in \citet{btfr}, a significant underestimation of the circular speeds of our sample is extremely unlikely. Further, since galaxies from several independent samples analyzed with independent techniques all seem to follow the trend in Figure~\ref{fig:VD_BTFR}, so the circular speeds of all the other galaxies would need to be underestimated in precisely the same way, which seems very unlikely. Finally, let us consider, ad absurdum, the following scenario. If we assume that all galaxies that deviate from the SPARC BTFR have wrong measurements, but they actually lie on it with $V_{\rm circ} \sim 80$~km~s$^{-1}$, then those galaxies should have higher surface brightness than dwarfs with $V_{\rm circ} \sim 20-40$~km~s$^{-1}$. So, if the trend in Figure~\ref{fig:VD_BTFR} is spurious, we should also find that galaxies which (apparently, due to wrong measurements) deviate from the BTFR have higher surface brightness than the dwarfs ($V_{\rm circ} \sim 20-40$~km~s$^{-1}$) in the BTFR, which is not observed. Based on this we are led to believe that the correlation in Figure~\ref{fig:VD_BTFR} is real, and it provides an extra parameter to explain the deviation from the canonical BTFR and its larger scatter at the low-velocity regime.

The vertical offset from the BTFR can also be seen as a progression in the baryon fraction of the galaxies: at fixed $V_{\rm circ}$, the more baryonic mass a galaxy has, the higher its baryon fraction is. This, coupled with our results above, implies that at low circular speeds $R_{\rm d}$ plays a role affecting the baryon fraction of galaxies (see Section~\ref{sec:failedfeedback} for more details).

Based on our discussion, we outline a possible interpretation of our results regarding the phenomenology of the BTFR. Perhaps, the SPARC BTFR holds at low circular speeds ($V_{\rm circ} \lesssim 50$~km~s$^{-1}$), but the distribution of galaxies in the $V_{\rm circ}-M_{\rm bar}$ plane may be more complex than a well-defined and tight relation as the one established at larger circular velocities, with dwarf galaxies showing baryon fractions above the one implied by the canonical BTFR but still below the cosmological limit. From our analysis, it looks very possible that the disc scale length is an important parameter regulating the deviation from the canonical BTFR.
A more extreme scenario would be one where the canonical BTFR breaks down at low circular speeds, being replaced by a 2D distribution. Given the selection biases and small statistics of the samples being analysed, we cannot discern among these options, and this should be addressed with more complete and representative samples.

\section{Discussion: the origin of gas-rich UDGs}
\label{sec:origins}
Using the kinematic information derived in the previous sections, here we discuss how our results compare with predictions from some of the main theories that have been proposed to explain the origin and properties of UDGs. 

\subsection{Brief comparison with NIHAO simulations: formation via feedback-driven outflows?}
\citet{dicintio} studied simulated dwarf galaxies from the Numerical Investigation of a Hundred Astrophysical Objects (NIHAO) simulations \citep{nihao}, and found a subset of them with properties similar to observed UDGs in isolation. They found that intermittent feedback episodes associated with bursty star formation histories modify the dark and luminous matter distribution, allowing dwarf galaxies to expand, as their baryons move to more external orbits (see also e.g., \citealt{navarro96,readgilmore,pontzen} or \citealt{readAD} for further considerations). Because of this, some of their simulated dwarf galaxies become larger, entering in the classification of UDGs. \citet{chan} reported similar results with the FIRE simulations (e.g., \citealt{FIRE}). We have observational evidence suggesting that our galaxies have low velocity dispersions and thus a low turbulence in the ISM. In principle, this seems at odds with models that require stellar feedback strong enough to modify the matter distribution. A detailed comparison between our observations and this kind of simulations it is beyond the scope of this paper. Yet, it is interesting to make some brief comments on some apparent similarities and discrepancies between the simulated NIHAO UDGs and our sample. 

By inspecting the optical scale lengths, we see that our largest galaxies have no counterparts among the NIHAO UDGs (their largest simulated UDG has $R_{\rm d} \approx 2$~kpc). In general, the mean values differ by a factor 2.5 (ours being larger), but strong selection effects are at play so this should be studied with a complete sample. The gas mass of our galaxies and NIHAO UDGs largely overlap, but our distribution has a sharp selection cut around $M\mathrm{_{HI}} < 10^{8.5}$~M$_\odot$. 

The UDGs formation mechanism proposed by \citet{dicintio} can also be contrasted with the observational results of \citet{btfr}, in particular the baryon fraction of the galaxies with respect to the cosmological mean and the inner amount of dark matter. \citet{dicintio} mention that their simulated UDGs show a correlation between their optical size and baryon fraction, with their largest UDG having a baryon fraction up to 50 per cent of the cosmological value, with a mean of 20 per cent for the whole sample. Our UDGs have $\approx 100$ per cent of the cosmological value. Nevertheless, as mentioned before, our galaxies have also larger scale lengths than the NIHAO UDGs, so one may wonder whether their higher baryon fraction is just a consequence of this. Extending our sample to include UDGs with smaller $R_{\rm e}$ may shed light on the connection between them and the simulated NIHAO UDGs. The inner dark matter content is a major discrepancy between our observations and the UDGs that the NIHAO simulation produces: our galaxies show very low dark matter fractions within their discs (measured within $\sim 2~R_{\rm e}$ on average), while \citealt{jiangUDGs} found that the NIHAO UDGs are centrally dark-matter dominated (measuring the dark-matter content within 1~$R_{\rm e}$). Related to this, \citet{dicintio} reported that their UDGs have dark matter concentration parameters typical of galaxies with similar halo masses. This does not seem to be the case in our sample: preliminary attempts of rotation curve decomposition of our UDGs show that if they inhabit `normal' NFW \citep{nfw} dark matter haloes (i.e. with a halo mass typical of galaxies with their stellar mass), their concentration parameters need to be extremely low (see also \citealt{sengupta}), far off expectations of canonical concentration-halo mass relations (e.g., \citealt{concentration}). This should be investigated further with data at higher spatial resolution, but it opens the exciting possibility of providing clues on the nature of dark matter itself (e.g. \citealt{yang2020}). Producing artificial data cubes of the NIHAO UDGs to explore their H\,{\sc i} kinematic parameters (like their position with respect to the BTFR), as well as obtaining SFR histories for our sample would also allow an interesting and conclusive comparison, although the latter has been proved to be challenging even for closer UDGs (e.g., \citealt{ruizlara, martinnavarro}). Stellar kinematics seems to be a promising tool as well \citep{nihaoxxiv}.

\subsection{High angular momentum}
\label{sec:angularmomentum}

Angular momentum is a fundamental quantity to understand the origin of high surface brightness and LSB galaxies (e.g., \citealt{dalcantondisks,dicintio2}), and it is usually studied via the so-called spin or $\lambda$--parameter for dark matter haloes (e.g., \citealt{mo,duttonvdb,aldo,posti}) or with the specific angular momentum--mass relation for the stellar component (e.g., \citealt{fall,rf12,fallandrom,posti2}). 

One of the main ideas to explain the large scale lengths and faint luminosities of UDGs is that they are dwarfs living in high spin (high-$\lambda$) dark matter haloes. This is supported by some semi-analytical models and hydrodynamical simulations, where the size of a galaxy is set by its $\lambda$, that seem to reproduce different observational properties of the (\textit{cluster}) UDG population like abundance, colours and sizes (e.g., \citealt{amorisco, rong, auriga}). Some other simulations, however, do not find anything atypical in the angular momentum content of UDGs (e.g., \citealt{dicintio, tremmel}).

In this Section we investigate the angular momentum content of our sample, looking separately at the specific angular momentum of gas and stars.

\subsubsection{H\,{\sc i} specific angular momentum}
Based on H\,{\sc i} observations, \citetalias{leisman} and \citet{spekkens} suggested that gas-rich UDGs could indeed have higher $\lambda$--parameter than other galaxies of similar mass. However, these results are derived from the relation given by \citet{hernandez} to estimate $\lambda$ from observations, which is highly assumption-dependent, as discussed in detail in \citet{duttonvdb}. In particular, our galaxies do not follow the same BTFR nor seem to have the same disc mass fraction as the galaxies used by \citet{hernandez} to calibrate their relation. Therefore, we decided not to estimate the $\lambda$--parameter in that way, and we emphasize that the calibration of \citet{hernandez} should be used with caution, as also mentioned in \citetalias{leisman}, \citet{spekkens} and \citet{sengupta}.

Unfortunately, we cannot robustly estimate the angular momentum of the gas component of our galaxies as we lack the resolution needed to determine the shape of the surface density profile (e.g., \citealt{obreschkow,kurapati}), so we can only make qualitative statements (see also \citealt{sengupta}). In this context, the fact that our gas-rich UDGs lie on the H\,{\sc i} mass--size relation \citep{gault} is useful, as it tells us that their H\,{\sc i} discs have normal sizes for their H\,{\sc i} mass. Additionally, we have shown that the gas rotates at velocities much lower than other galaxies with the same H\,{\sc i} mass. Together, these results suggest that our UDGs have low-to-normal gas specific angular momenta compared with galaxies of similar H\,{\sc i} mass.

\subsubsection{Stellar specific angular momentum}

As mentioned before, the stellar specific angular momentum--mass relation (sometimes called `Fall' relation, see \citealt{fall,rf12}) is often used as a more direct way to study the angular momentum of galaxies. 
To compute this relation, high-resolution (stellar) rotation curves and stellar surface density profiles are needed. However, it is common (see discussion in \citealt{rf12,rizzo}) to adopt the approximation
\begin{equation}
    j_\star = 2~R_{\rm d}~V_{\rm rot,\star}~,
    \label{eq:jstar}
\end{equation}
where $j_\star$ is the stellar specific angular momentum, $R_{\rm d}$ the optical disc scale length and $V_{\rm rot,\star}$ the stellar rotation velocity. This approximation has been proved to work very well, and it is valid for galaxies with exponential light profiles and flat rotation curves. Thus, to use this simplified version to compute $j_\star$ for our sample, we have to assume flat rotation curves, which seems at least tentatively supported by our data, and exponential profiles, that describe reasonably well the stellar profile of our galaxies (see \citealt{gault}).

As the rotation velocity of the stars is needed, the next step is assuming that their rotation can be inferred from the circular velocity of the galaxies, by means of the stellar asymmetric drift correction ($V_{\rm AD,\star}$), via the equation
\begin{equation}
V_{\rm rot,\star}^2 = V_{\rm circ}^2 - V_{\rm AD,\star}^2~.
\end{equation}

\noindent
To compute $V_{\rm AD,\star}$, we follow the approach described in \citet{posti2}, using the equation:
\begin{equation}
\label{eq:AD}
V^2_{\rm AD,\star} = \upsigma^2_{0,z} \dfrac{3R}{2R_{\rm d}} e^{-R/R_{\rm d}}~,
\end{equation}
with $\upsigma_{0,z}$ the vertical stellar velocity dispersion\footnote{Eq.~\ref{eq:AD} assumes isotropy, i.e., $\upsigma_R = \upsigma_z$. However, as explored by \citet{posti2}, the difference between this or assuming extreme anistropic profiles is rather small, of less than 10 per cent.}. For simplicity we use only the outermost point of the rotation curve, so effectively $R = R_{\rm out}$. As discussed by \citet{posti2}, the value of $\upsigma_{0,z}$ depends on the central surface brightness \citep{martinsson}, and for our sample it is about 5~km~s$^{-1}$.

\begin{figure}
    \centering
    \includegraphics[scale=0.51]{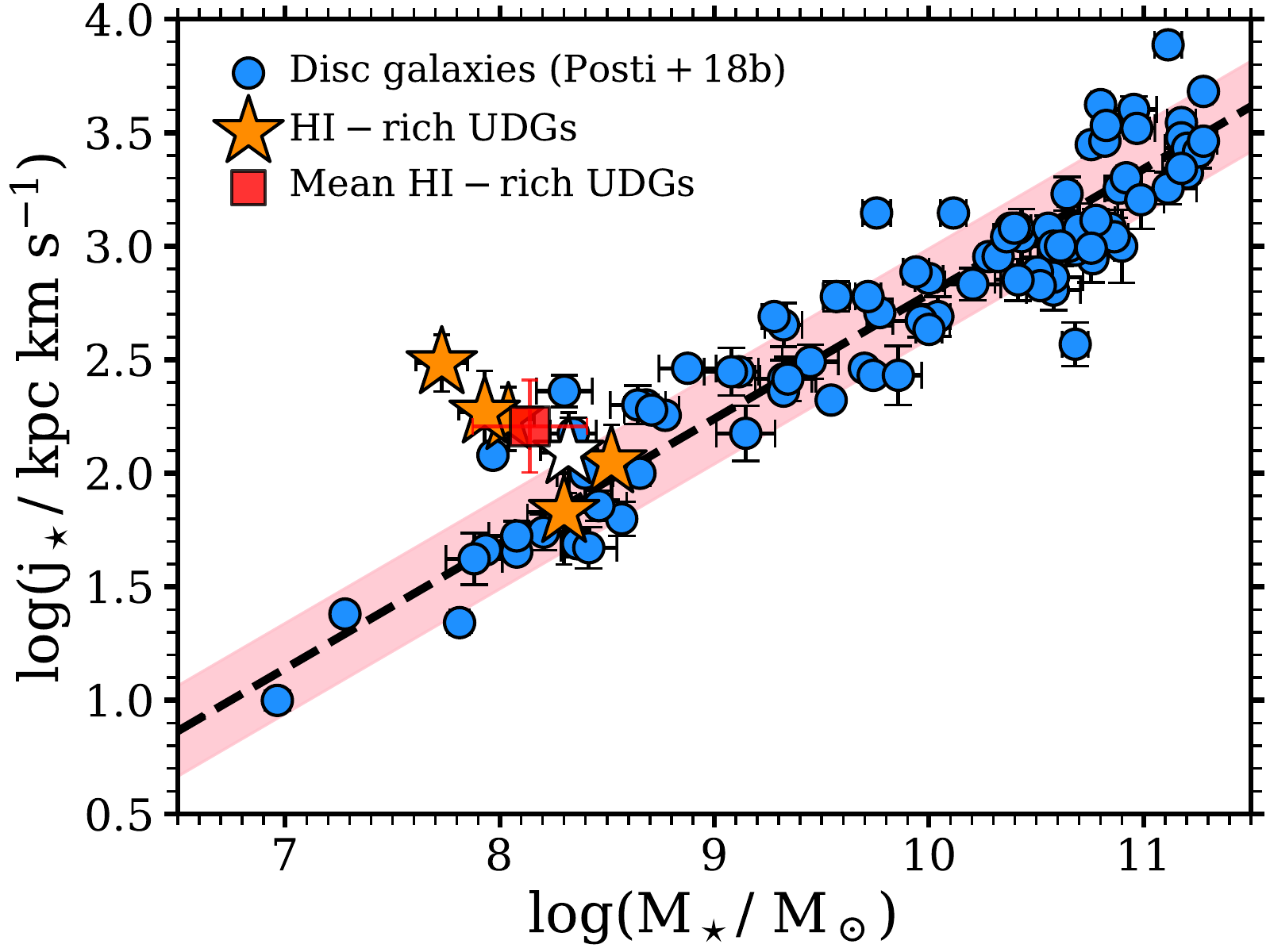}
    \caption{Stellar specific angular momentum--mass relation. Orange stars show our UDGs (AGC~749290 is in white as in Figure~\ref{fig:BTFRall}) and the red square their mean position. Blue circles show the sample analyzed by \citet{posti2}, while the black dashed line and the pink band are their best-fit relation and its 1$\upsigma$ scatter, respectively.}
    \label{fig:jstar}
\end{figure}

The results of this exercise are shown in Figure~\ref{fig:jstar}, where we plot our galaxies in the $M_\star$ vs. $j_\star$ plane. We compare our UDGs with the galaxies studied in \citet{posti2}, showing also the best-fit relation (dashed line) and scatter (pink band) that they obtain. We stress here that the assumptions that we have made in our analysis are the same as in \citet{posti2}, making the comparison in Figure~\ref{fig:jstar} as fair as our data allow.

Three of our galaxies (AGC~114905, AGC~248945 and AGC~749290) lie within the 1$\upsigma$ scatter of the relation. AGC~219533 and AGC~334315 have a $j_\star$ about 3--4 times larger than the best-fit line, although the observational scatter is relatively large at those values of $M_*$. The outlier with the highest $j_\star$ is AGC~122966, which has both the largest optical disc scale length and highest rotation velocity of our sample, resulting in a $j_\star$ about 9 times larger than expected. Note, however, that the Fall relation is not well-constrained at $M_\star < 10^8$~M$_\odot$. 
A caveat to bear in mind regarding AGC~122966 is that it has the lowest surface brightness of our sample (see Table~1 in \citealt{btfr}), so its scale length is relatively more uncertain than for the other UDGs. 
The mean (median) ratio between the measured and expected $j_\star$ of our galaxies is 3.3 (2.5). These numbers are of course dependent on the several assumptions we have made, and need to be confirmed with better data and more accurate calculations (for instance by obtaining high-resolution stellar rotation curves to formally compute $j_\star$ instead of using the approximation of Eq.~\ref{eq:jstar} with Eq.~\ref{eq:AD}). Yet, our simple analysis indicates that, as a population, our gas-rich UDGs may have a $j_\star$ a factor $\approx 3$ higher than the expectations for dwarf irregular galaxies with similar $M_\star$, as shown graphically with in Figure~\ref{fig:jstar}, where the mean value for our sample is indicated in red. This larger $j_\star$ may help explaining why UDGs have more extended optical disc scale length/effective radius than other dwarf irregulars.

The H\,{\sc i} component of our galaxies is both more massive and more extended, so one may speculate that its specific angular momentum is likely to be more representative of the spin of the dark matter halo. If this is the case, our UDGs would be galaxies that inhabit dark haloes with normal-to-low $\lambda$ but with higher-than-average $j_\star$, meaning that they would be galaxies with a higher-than-average `retained'  fraction of angular momentum ($j_\star / j_{\rm halo}$), as suggested by \citet{posti}.

\subsection{`Failed' Milky Way galaxies}
\label{sec:failed}
Another mechanism proposed to explain the nature of UDGs is that they could be `failed' Milky Way-like galaxies, with massive dark matter haloes that for different reasons (e.g., strong supernova feedback or gas stripping) failed at converting their gas into stars \citep{vandokkum16}. This idea is mainly motivated by the high velocity dispersions of globular cluster observed around a few UDGs (e.g., \citealt{vandokkum16,toloba}) and by the large effective radius of UDGs (but see \citealt{chamba}). However, several other studies, both observational and theoretical, show that most UDGs should reside in dwarf galaxy-sized dark matter haloes (e.g., \citealt{beasleyytrujillo, rong, amoriscohaloes, ruizlara, kovacs, prolegcs, tremmel}).

A dark matter NFW halo of virial mass 10$^{12}$~M$_\odot$, following a standard concentration-mass relation, is expected to have a maximum circular speed around 170~km~s$^{-1}$, with a value of about 120~km~s$^{-1}$ at our typical $R_{\rm out}$, much larger than the velocities observed in our sample. Thus, our data can safely exclude the `failed' Milky Way scenario as an origin for our gas-rich UDGs.

Actually, the fact that our galaxies have a baryon fraction close to the cosmological mean (see Figure~\ref{fig:BTFRall}) suggests that they may reside in $\sim 10^{10}$~M$_\odot$ dark mater haloes\footnote{This comes from the fact that the cosmological baryon fraction $f_{\rm bar} = M_{\rm bar}/M_{\rm DM} \approx 0.16$, so a galaxy with a baryon fraction equal to $f_{\rm bar}$ should reside in a dark matter halo with $M_{\rm DM} = M_{\rm bar}/f_{\rm bar} \approx 6 M_{\rm bar}$.}, although this should be confirmed with a detailed mass decomposition and will be the subject of future work.

\subsection{Ancient tidal dwarfs}
Tidal dwarf galaxies (TDGs) are self-gravitating systems formed from the collapsed tidal debris of interacting galaxies. Because of this, one expects to find them inhabiting the chaotic environments near their progenitors, even after some Gyr (e.g., \citealt{duc,lelliTD}). In fact, TDGs in the nearby universe are usually found within 15~$R_{\rm e,p}$ of their progenitors ($R_{\rm e,p}$ being the effective radius of the progenitor, \citealt{kaviraj}). As they form from pre-enriched material they are expected to show high metallicities (e.g., \citealt{ducmirabel}, but see also \citealt{hunterTD}), and due to their weak gravitational potential they should be free of dark matter (see \citealt{hunterTD,braine,lelliTD} and references therein).

From an observational point of view, TDGs present some properties similar to those in our sample of UDGs. For instance, it has been argued that they are comparable in terms of effective radius and surface brightness \citep{duc}. Perhaps more intriguingly, they also share dynamical properties: they lie off the BTFR and show dark matter fractions close to zero within their H\,{\sc i} discs \citep{lelliTD}. Given all this, it is pertinent to ask whether or not our gas-rich UDGs may be TDGs. In this Section we explain why we consider this scenario unlikely.

The strongest evidence against a tidal origin is the environment of our UDGs. They cannot be \textit{young} tidal dwarfs given their totally different environments: they are isolated with the mean distances to their nearest, second-nearest and third-nearest neighbors being 1~Mpc, 1.4~Mpc and 1.7~Mpc, respectively. This is completely different from the expected progenitor-TDG separation of 15~$R_{\rm e,p}$, which would typically be around 100~kpc. It would be required that \textit{all} our galaxies formed as TDGs some Gyr ago, and the separation between all of them and the other interacting galaxy increased up to at least 1~Mpc, which seems unlikely (see also the text in Section~\ref{sec:circularspeeds}). 

One may also argue that we do not find remaining signs of tidal interactions around our galaxies, but this could be just because detecting such interactions is hard (e.g., \citealt{holwerda,muller}). Perhaps more importantly, from cosmological simulations one would expect TDGs to have smaller sizes than typical dwarfs (see the discussion in \citealt{tdgs_illustris}), while UDGs are exactly the opposite, a population of galaxies with much larger optical radii than other dwarfs; \citet{duc}, however, argue that some old TDGs are larger than normal dwarfs.

Currently, our interpretation is that our galaxies live in $\sim$~10$^{10}$~M$_\odot$ dark mater haloes and the dark-matter deficiency is restricted to the extent of the H\,{\sc i} disc. However, our current data cannot unambiguously distinguish between this scenario and one where our UDGs have very little, if any, dark matter in their whole extension, as expected in TDGs. The mass decomposition in our galaxies would then conclusively tell us if in fact there is room to accommodate (low-concentration) 10$^{10}$~M$_\odot$ dark mater haloes or if their lack of dark matter is analogous to that in TDGs.

\subsection{Weak feedback producing little mass losses}
\label{sec:failedfeedback}

Different cosmological hydrodynamical simulations predict large mass loading factors in dwarf galaxies (e.g., \citealt{vogelsberger, ford}), although other theoretical studies show that big mass losses do not necessarily take place (see for instance \citealt{strickland, dalcanton2007,emerick,romano}). Supporting this latter scenario, there is recent observational evidence (\citealt{mcquinn}, see also \citealt{lelli2014_2}) that the mass loading factors in dwarf galaxies are indeed relatively small, as often the outflows do not reach the virial radii of the galaxies and are kept inside their haloes, available for the regular baryon cycle. These results suggest that some dwarfs may have baryon fractions larger than expected, qualitatively in agreement with our gas-rich UDGs with `no missing baryons'. 

In this work, along with \citet{btfr}, we suggest that a scenario where feedback processes in our UDGs have been relatively weak and inefficient in ejecting gas out of their virial radii could explain their quiescent ISM (inferred from the velocity dispersion) and high baryon fractions (as derived from the BTFR). In Figure~\ref{fig:VD_BTFR}, we found a significant trend for low-mass galaxies (15~km~s$^{-1}< V_{\rm circ} < 45$~km~s$^{-1}$) where they deviate more from the canonical BTFR, towards higher masses (and thus have a larger baryon fraction), if they have large disc scale lengths. As most of these galaxies have normal or low SFRs with respect to their stellar masses (e.g., \citealt{teich}; \citetalias{leisman}, but more robust measurements would be desirable), their large disc scale lengths then imply that they have lower-than-average (about an order of magnitude) SFR surface densities, and we speculate that this affects their capability to drive outflows powerful enough to eject baryons out of their virial radius, and thus allowing the galaxies to retain a higher-than-average baryon fraction. Recently, using high-resolution hydrodynamical simulations, \citet{romano} have shown, for an ultra faint dwarf galaxy, that this indeed may be the case: the gas removal of a galaxy depends on the distribution of supernovae explosions and thus in the distribution of star formation. Those authors find that the more evenly spread or `diluted' the distribution of OB associations is, the higher is the gas fraction that the galaxy keeps, in agreement with our interpretation.

\section{Conclusions}
\label{sec:conclusions}

In this work we present the 3D kinematic models of six dwarf gas-rich ultra-diffuse galaxies (UDGs). By analysing VLA and WSRT 21-cm observations with the software $\mathrm{^{3D}}$Barolo, we derive reliable measurements of the circular speed and gas velocity dispersion of our sample galaxies. Our models have been used by \citet{btfr} to show that the galaxies lie significantly above the baryonic Tully-Fisher relation.\\

\noindent
Our main findings can be summarised as follows:
\begin{itemize}
    \item We have shown that the kinematic models are robust (Figures~\ref{fig:114905}-\ref{fig:inclinations}) and our galaxies have circular speeds of $20-40$~km~s$^{-1}$.
    \item Our UDGs exhibit low gas velocity dispersions, lower than observed in most dwarf irregular galaxies before. Their H\,{\sc i} layers have the typical thicknesses observed in other dwarfs and disc galaxies, and gas turbulence appears to be fed by supernovae with efficiencies of just a few percent.
    \item We have reviewed the canonical baryonic Tully-Fisher relation, showing that these gas-rich UDGs are likely not the only outliers, although they may represent the most extreme cases (Figure~\ref{fig:BTFRall}). 
    \item At circular speeds below $\approx$~45~km~s$^{-1}$ the vertical deviation from the canonical baryonic Tully-Fisher relation correlates with the disc scale length of the galaxies (Figure~\ref{fig:VD_BTFR}).
    \item The low velocity dispersions observed in our sample seem at odds with models where UDGs originate from feedback-driven outflows. Our galaxies tend to have larger scale lengths than the simulated NIHAO UDGs and to have higher baryon fractions, but they share some other structural properties. The most important discrepancy is that, unlike NIHAO simulated UDGs, our galaxies have little dark matter in the inner regions (within $\sim 2 R_{\rm e}$).
    \item We find indications that the gas specific angular momentum of our sample is similar or slightly lower than that in other galaxies of similar H\,{\sc i} mass. However, the specific angular momentum of the stellar component may be (a factor $\approx 3$) higher-than-average at given $M_\star$ (Figure~\ref{fig:jstar}). This can help to explain the large optical scale length of UDGs.
   \item The measured low circular speeds rule out the possible origin as `failed' Milky Way galaxies for our UDGs.
    \item Fully testing the idea that all our six galaxies are old tidal dwarf galaxies is not possible with our observations, but their isolation seems to go against this possibility.
    \item To explain the high baryon fractions and low turbulence in the discs, \citet{btfr} have suggested that these galaxies experienced weak feedback, allowing them to retain all of their baryons. Here we have shown that this idea is consistent with our findings: the larger the optical disc scale length of dwarf galaxies is, the more they depart from the canonical baryonic Tully-Fisher relation towards higher baryonic masses. Their extended optical sizes coupled with normal star formation rates result in very low star formation rate surface densities, impacting their capability to lose mass via outflows.
\end{itemize}{}

\section*{Acknowledgements}
We thank an anonymous referee for the comments provided on our manuscript. We would like to thank Arianna di Cintio, Vladimir Avila-Reese, Anastasia Ponomareva, Nushkia Chamba and Francesco Calura for fruitful discussions regarding our data and results. We also acknowledge the work done by Michael Battipaglia, Elizabeth McAllan and Hannah J. Pagel on acquisition and reduction of data used in this work. P.E.M.P. thanks Giuliano Iorio and Salvador Cardona for different comments and clarifications regarding LITTLE THINGS and NIHAO data, respectively. 
P.E.M.P., F.F. and T.O. are supported by the Netherlands Research School for Astronomy (Nederlandse Onderzoekschool voor Astronomie, NOVA), Phase-5 research programme Network 1, Project 10.1.5.6. K.A.O. received support from VICI grant 016.130.338 of the Netherlands Organization for Scientific Research (NWO). E.A.K.A. is supported by the WISE research programme, which is financed by NWO. G.P. acknowledges support by the Swiss National Science Foundation grant PP00P2\_163824. L.P. acknowledges support from the Centre National d'\'{E}tudes Spatiales (CNES). M.P.H. acknowledges support from NSF/AST-1714828 and the Brinson Foundation. K.L.R. is supported by NSF grant AST-1615483.

This work uses observations from the VLA and the WSRT. The VLA is operated by the National Radio Astronomy Observatory, a facility of the National Science Foundation operated under cooperative agreement by Associated Universities, Inc. The WSRT is operated by the ASTRON (Netherlands Institute for Radio Astronomy) with support from NWO. Our work is also based in part on observations at Kitt Peak National Observatory, National Optical Astronomy Observatory, which is operated by the Association of Universities for Research in Astronomy (AURA) under a cooperative agreement with the National Science Foundation.

We have made extensive use of SIMBAD and ADS services, as well as of the Python packages NumPy \citep{numpy}, Matplotlib \citep{hunter}, SciPy \citep{scipy} and Astropy \citep{astropy}, and the tool TOPCAT \citep{topcat}, for which we are thankful.






\begin{thebibliography}{199}
  
  
 \bibitem[Adams \& Oosterloo(2018)]{leot} Adams, E. A. K. \& Oosterloo, T. 2018, A\&A, 612, 26
  
    \bibitem[Amorisco(2018)]{amoriscohaloes} Amorisco, N. C. 2018, MNRAS, 475, L116
  
  \bibitem[Amorisco \& Loeb(2016)]{amorisco} Amorisco, N.C., \& Loeb, A. 2016, MNRAS, 459, 51
  
\bibitem[Astropy Collaboration et al.(2013)]{astropy} Astropy Collaboration et al. 2013, A\&A, 558, A3
  
  \bibitem[Avila-Reese et al.(2008)]{vladimir} Avila-Reese V., Zavala J., Firmani C., Hern\'andez-Toledo H. M.,2008, AJ, 136, 1340
  
  \bibitem[Bacchini et al.(2019)]{ceci} Bacchini, C., Fraternali, F., Iorio, G., Pezzulli, G., 2019, A\&A, 622, 64
  
  \bibitem[Banerjee et al.(2011)]{banerjee} Banerjee, A., Jog, C. J., Brinks, E., Bagetakos, I., 2011, MNRAS, 415, 687
  
  
  
  
        \bibitem[Beasley \& Trujillo(2016)]{beasleyytrujillo} Beasley, M. A., Trujillo, I., ApJ, 830, 26
  
  \bibitem[Bennet et al.(2018)]{bennet} Bennet P., Sand D. J., Zaritsky D., Crnojevi\'c D., Spekkens K., Karunakaran A., 2018, ApJ, 866, L11
  
 \bibitem[Bosma(1978)]{bosma} Bosma A., 1978, PhD thesis, University of Groningen
  
  \bibitem[Braine et al.(2001)]{braine} Braine, J., Duc, P.-A., Lisenfeld, U., et al. 2001, A\&A, 378, 51
  
  
  \bibitem[Cardona-Barrero et al.(2020)]{nihaoxxiv}  Cardona-Barrero, S., Di Cintio, A., Brook, C. B. A., Ruiz-Lara, T., et al. 2020, arXiv:2004.09535
  
  \bibitem[Cimatti, Fraternali \& Nipoti(2019)]{bookFilippo} Cimatti, A., Fraternali, F., \& Nipoti, C. 2019. Introduction to Galaxy Formation and Evolution: From Primordial Gas to Present-Day Galaxies. Cambridge: Cambridge University Press.
  
  \bibitem[Chamba, Trujillo \& Knapen(2020)]{chamba}  Chamba, N., Trujillo, I., \& Knapen, J. H. 2020, A\&A, 633L, 3 
  
  \bibitem[Chan et al.(2018)]{chan} Chan T. K. et al. 2018,  MNRAS, 478, 906
  
  


 
 \bibitem[Dalcanton(2007)]{dalcanton2007} Dalcanton, J. 2007, ApJ, 658, 941
 
    \bibitem[Dalcanton et al.(1997b)]{dalcantondisks} Dalcanton, J. et al. 1997b, ApJ, 482, 659
      
      

      
      
      \bibitem[Di Teodoro \& Fraternali(2014)]{enrico14} Di Teodoro E. M. \& Fraternali F., 2014, A\&A, 567, A68
      
      \bibitem[Di Teodoro \& Fraternali(2015)]{barolo} Di Teodoro, E. M. \& Fraternali, F., 2015, MNRAS, 451, 3021
      
  \bibitem[Di Teodoro et al.(2016)]{enrico2} Di Teodoro, E. M., Fraternali, F., \& Miller, S. H. 2016, A\&A, 594, A77
      
    \bibitem[Di Cintio et al.(2017)]{dicintio} Di Cintio, A., Brook, C. B., Dutton, A. A., et al. 2017, MNRAS, 466, L1

    \bibitem[Di Cintio et al.(2019)]{dicintio2} Di Cintio, A. et al. 2019, MNRAS, 486, 2535

 \bibitem[de Blok(1997)]{deblok} de Blok, W. J. G. 1997, PhD thesis, University of Groningen
 
 \bibitem[de Blok \& Walter(2000)]{deblok2000} de Blok W. J. G., Walter F., 2000, ApJL, 537, L95

\bibitem[Duc \& Mirabel(1998)]{ducmirabel} Duc, P.-A., \& Mirabel, I. F. 1998, A\&A, 333, 81

\bibitem[Duc et al.(2014)]{duc} Duc, P.-A., Paudel, S., McDermid, R. M., et al. 2014, MNRAS, 440, 1458


\bibitem[Dutton \& van den Bosch(2012)]{duttonvdb} Dutton, A. A. \& van den Bosch, F. C. 2012, MNRAS, 421, 608




\bibitem[Emerick et al.(2018)]{emerick} Emerick A., Bryan G. L., Mac Low M.-M., 2018, ApJ, 865, L22

\bibitem[Falc\'on-Barroso et al.(2011)]{falconbarroso} Falc\'on-Barroso J., Balcells M., Peletier R. F., Vazdekis A., 2003, A\&A, 405, 455

\bibitem[Fall(1983)]{fall} Fall S. M., 1983, in Athanassoula E., ed., IAU Symposium Vol.100, Internal Kinematics and Dynamics of Galaxies. pp 391–398

\bibitem[Fall \& Romanowsky(2018)]{fallandrom} Fall S.M., Romanowsky A.J.,2018, ApJ, 868, 133

\bibitem[Fattahi et al.(2016)]{apostle2} Fattahi, A., Navarro, J. F., Sawala, T., et al. 2016, MNRAS,457, 844

\bibitem[Field(1965)]{field} Field, G. B. 1965, ApJ, 142, 531.

\bibitem[Fierlinger et al.(2016)]{fierlinger} Fierlinger K. M., Burkert A., Ntormousi E., Fierlinger P., Schartmann M., Ballone A., Krause M. G. H., Diehl R., 2016, MNRAS, 456, 710

\bibitem[Ford et al.(2014)]{ford} Ford, A. B., Dav\'e, R., Oppenheimer, B. D., et al. 2014, MNRAS, 444, 1260

\bibitem[Fraternali et al.(2002)]{ff02} Fraternali, F., van Moorsel, G., Sancisi, R., \& Oosterloo, T. 2002, AJ, 123, 3124

\bibitem[Gault et al.(submitted)]{gault} Gault, L., et al., submitted to ApJ

\bibitem[Geha et al.(2006)]{geha} Geha, M., Blanton, M. R., Masjedi, M., \& West, A. A. 2006, ApJ, 653, 240

\bibitem[Giovanelli et al.(2005)]{alfalfa} Giovanelli, R., Haynes, M. P., Kent, B. R., et al. 2005, AJ, 130, 2598

\bibitem[Gooch(2006)]{karma} Gooch R., 2006, Karma Users Manual. Australia Telescope National Facility


\bibitem[Greco et al.(2018)]{greco} Greco, Johnny P., et al. 2018, ApJ, 857, 104

\bibitem[Guo et al.(2019)]{guo} Guo, Q. et al. 2019, arXiv:1908.00046

\bibitem[Harbeck et al.(2014)]{harbeck} Harbeck, D. R., Boroson, T., Lesser, M., et al. 2014, Proc. SPIE, 9147, 9147E

\bibitem[Haslbauer et al.(2019)]{tdgs_illustris} Haslbauer, M., et al. 2019, A\&A, 626, 47

\bibitem[Haurberg et al.(2015)]{haurberg} Haurberg, N. C., Salzer, J. J., Cannon, J. M., \& Marshall, M. V. 2015, ApJ, 800, 121

\bibitem[He et al.(2019)]{he} He, M. et al. 2019, ApJ, 880, 30

\bibitem[Heiles \& Troland(2003)]{heiles} Heiles, C., \& Troland, T. H. 2003, ApJ, 586, 1067

\bibitem[Hernandez et al.(2007)]{hernandez} Hernandez, X. et al. 2007, MNRAS, 375, 163

\bibitem[Herrmann et al.(2016)]{herrmann} Herrmann K. A., Hunter D. A., Zhang H.-X., Elmegreen B. G., 2016, AJ, 152, 17

\bibitem[Holwerda et al.(2011)]{holwerda} Holwerda B. W., Pirzkal N., Cox T. J., de Blok W. J. G., Weniger J., Bouchard
A., Blyth S.-L., van der Heyden K. J., 2011, MNRAS, 416, 2426

\bibitem[Hopkins et al.(2018)]{FIRE} Hopkins P. F., et al., 2018, MNRAS, 480, 800

  \bibitem[Hunter(2007)]{hunter} Hunter J. D., 2007, Computing In Science \& Engineering, 9, 90
  
  \bibitem[Hunter \& Elmegreen(2006)]{hunterLT} Hunter, D. A. \& Elmegreen, B. G. 2006, ApJS, 162, 49
  
  \bibitem[Hunter et al.(2000)]{hunterTD} Hunter D. A., Hunsberger S. D., Roye E. W., 2000, ApJ, 542, 137
  
 \bibitem[Impey et al.(1988)]{impey} Impey C., Bothun G., Malin D., 1988, ApJ, 330, 634
  
  
  \bibitem[Iorio et al.(2017)]{iorio} Iorio G., Fraternali F., Nipoti C., Di Teodoro E., Read J. I., Battaglia G., 2017, MNRAS, 466, 415
  
   \bibitem[Janowiecki et al.(2019)]{jano} Janowiecki, S. et al. 2019, MNRAS, DOI: 10.1093/mnras/stz1868

  
  \bibitem[Jiang(2019b)]{jiangUDGs} Jiang, F. et al. 2019, MNRAS, 487, 5272
  
  \bibitem[Jones et al.(2001)]{scipy} Jones, E. et al., 2001. SciPy: Open source scientific tools for Python, Available at: http://www.scipy.org/

  
    \bibitem[Jones et al.(2018)]{jones} Jones, M. G., et al. 2018, A\&A, 614, 21

\bibitem[Kaviraj et al.(2012)]{kaviraj} Kaviraj, S. et al. 2012, MNRAS, 419, 70 



\bibitem[Kirby et al.(2012)]{kirby12} Kirby, E.M., Koribalski, B.S., Jerjen, H., L\'opez-S\'anchez, A.R. 2012, MNRAS 420, 2924

\bibitem[Kovacs et al.(2019)]{kovacs} Kovacs, O. E., et al. 2019, ApJ, 879, 12
  
  \bibitem[Kurapati et al.(2018)]{kurapati} Kurapati, S., Chengalur, J. N., Pustilnik, S., \& Kamphuis, P. 2018, MNRAS, 479,228
  
  


\bibitem[Lee et al.(2020)]{lee2020} Lee, J. H., Kang, J., Lee, M. G., Jang, I. S. 2020, to appear in ApJ, arXiv:2004.01340

    \bibitem[Leisman et al.(2017)]{leisman} Leisman L., et al. 2017, ApJ, 842, 13
    
\bibitem[Lelli et al.(2016a)]{sparc} Lelli F., McGaugh S. S., Schombert J. M., 2016a, AJ 152, 157

\bibitem[Lelli et al.(2016b)]{lelli2016} Lelli, F., McGaugh, S. S., Schombert, J. M., 2016b, ApJ, 816, 14

\bibitem[Lelli et al.(2015)]{lelliTD} Lelli, F. et al. 2015, A\&A, 584, A113


\bibitem[Lelli et al.(2014a)]{lelli2014_2} Lelli, F., Fraternali, F., \& Verheijen, M. 2014, A\&A, 563, A27

 \bibitem[Lelli et al.(2014b)]{lelli2014} Lelli, F., Verheijen  M.  A.  W., Fraternali, F. 2014, A\&A, 566, 71
 
 \bibitem[Leroy et al.(2009)]{leroy} Leroy, A. K., Walter, F., Bigiel, F., et al. 2009, AJ, 137, 4670

\bibitem[Liao et al.(2019)]{auriga} Liao, S. et al. 2019, MNRAS, 490, 5182 

\bibitem[Ludlow et al.(2014)]{concentration} Ludlow A. D. et al., 2014, MNRAS, 441,378

\bibitem[Mancera Pi\~na et al.(2019a)]{paperII} Mancera Pi\~na, P. E., Peletier, R.F., Aguerri, J. A. L., Venhola, A., Trager, S., Choque Challapa, N., 2019a, MNRAS, 485, 1036

\bibitem[Mancera Pi\~na et al.(2019b)]{btfr} Mancera Pi\~na, P. E., Fraternali, F., Adams, B., Marasco, A., Oosterloo, T., et al. 2019b, ApJL, 883, L33

    \bibitem[Mancera Pi\~na et al.(2018)]{paperI} Mancera Pi\~na, P. E., Peletier, R. F., Aguerri, J.A.L., Venhola, A., Trager, S., Choque Challapa, N., 2018, MNRAS, 481, 4381

\bibitem[Mac Low \& Klessen(2004)]{maclow} Mac Low M.-M., Klessen R. S., 2004, Reviews of Modern Physics, 76, 125

\bibitem[Marasco \& Fraternali(2011)]{marascoFF} Marasco A., Fraternali F., 2011, A\&A, 525, A134


\bibitem[Martinsson et al.(2013)]{martinsson} Martinsson, T. P. K., Verheijen, M. A. W., Westfall, K. B., et al. 2013, A\&A, 557,
A130

\bibitem[Mart\'in-Navarro et al.(2019)]{martinnavarro} Mart\'in-Navarro, I., et al. 2019, MNRAS, 484, 3425

\bibitem[Masters(2005)]{masters} Masters, K., 2005, PhD thesis, Cornell University




\bibitem[McGaugh \& de Blok(1998)]{mcgaugh98} McGaugh S. S. \& de Blok W. J. G., 1998, ApJ, 499, 41

\bibitem[McGaugh et al.(2000)]{mcgaugh1} McGaugh, S. S., Schombert, J. M.; Bothun, G. D.; de Blok, W. J. G., 2000, ApJ, 533L, 99

 \bibitem[McNichols et al.(2016)]{shield} McNichols, A. T., et al. 2016, ApJ, 832, 89
 
 \bibitem[McQuinn et al.(2019)]{mcquinn} McQuinn, K. B. W., van Zee, L., Skillman, E. D. 2019, arxiv:1910.04167


 \bibitem[Mo et al.(1988)]{mo} Mo, H. J., Mao S., White S. D. M., 1998, MNRAS, 295, 319

 \bibitem[M\"uller et al.(2019)]{muller} M\"uller, O., Rich, R. M., Rom\'an, J. 2019, A\&A, 624, 6

\bibitem[Navarro, Eke \& Frenk(1996)]{navarro96} Navarro J. F., Eke V. R., Frenk C. S., 1996, MNRAS, 283, L72

\bibitem[Navarro, Frenk \& White(1996)]{nfw} Navarro J. F., Frenk C. S. \& White S. D. M., 1996, ApJ, 462, 563

 \bibitem[Noordermeer(2006)]{noordermeer} Noordermeer, E. 2006, PhD thesis, University of Groningen
 

 
 \bibitem[Obreschkow \& Glazebrook(2014)]{obreschkow} Obreschkow D., Glazebrook K., 2014, ApJ, 784, 26



\bibitem[Oh et al.(2015)]{oh15} Oh S.-H. et al., 2015, AJ, 149, 180

  \bibitem[Oliphant(2006)]{numpy} Oliphant, T. 2006, A Guide to NumPy, Trelgol Publishing USA
  
   \bibitem[Oman et al.(2015)]{kyle2015} Oman, K., et al. 2015, MNRAS, 452, 3650
  
  \bibitem[Oman et al.(2016)]{kyle2016} Oman, K., Navarro, J. F., Sales, L. V., et al. 2016, MNRAS, 460, 3610
  
  \bibitem[Oman et al.(2019)]{kyle2019} Oman, K., et al. 2019, MNRAS, 482, 821O
  
  \bibitem[Oosterloo et al.(2007)]{oosterloo2007} Oosterloo, T., Fraternali, F., \& Sancisi, R. 2007, AJ, 134, 1019
  
     
     \bibitem[Ponomareva et al.(2018)]{anastasia2} Ponomareva, A., Verheijen, M. A. W.., Papastergis, E., Bosma, A., Peletier, R. F. 2018, MNRAS, 474, 4366
     
          \bibitem[Ponomareva et al.(2017)]{anastasia} Ponomareva, A.,  Verheijen  M.  A.  W.,  Peletier  R.  F.,  Bosma A., 2017, MNRAS, 469, 2387
     
     \bibitem[Pontzen \& Governato(2012)]{pontzen} Pontzen A., Governato F., 2012, MNRAS, 421, 3464
     
 \bibitem[Posti et al.(2018b)]{posti2} Posti, L., Fraternali, F., Di Teodoro, E. M., Pezzulli, G., 2018b, A\&A, 612, L6
 
 \bibitem[Posti et al.(2019a)]{postinonmissing} Posti, L., Fraternali, F., Marasco, A., 2019, A\&A, 626, 56
 
 \bibitem[Posti et al.(2018a)]{posti} Posti L., Pezzulli G., Fraternali F., Di Teodoro E. M., 2018a, MNRAS, 475, 232

\bibitem[Prole et al.(2019a)]{prolegcs} Prole D. J., et al., 2019, MNRAS, 484, 486

\bibitem[Prole et al.(2019b)]{prole} Prole, D. et al. 2019, MNRAS, 488, 2143

\bibitem[Read \& Gilmore(2005)]{readgilmore} Read, J. I.; Gilmore, G. 2005 MNRAS, 356, 107 

\bibitem[Read et al.(2016)]{readAD} Read, J. I., Iorio, G., Agertz, O., \& Fraternali, F. 2016, MNRAS, 462, 3628


\bibitem[Read et al.(2017)]{read} Read, J. I., Iorio, G., Agertz, O., \& Fraternali, F. 2017, MNRAS, 467, 2019


    
    \bibitem[Rizzo et al.(2018)]{rizzo} Rizzo, F., Fraternali, F., Iorio, G. 2018, MNRAS, 476, 2137
    
    \bibitem[Rodr\'iguez-Puebla et al.(2016)]{aldo} Rodr\'iguez-Puebla, A., Behroozi, P., Primack, J., et al. 2016, MNRAS, 462, 893
    
    \bibitem[Rom\'an et al.(2019)]{roman3} Rom\'an, J., et al. 2019, MNRAS, 486, 823
    
        \bibitem[Rom\'an \& Trujillo(2017a)]{roman} Rom\'an, J., \& Trujillo, I. 2017, MNRAS, 468, 703

    \bibitem[Rom\'an \& Trujillo(2017b)]{roman2} Rom\'an, J., \& Trujillo, I. 2017, MNRAS, 468, 4039
    
    \bibitem[Romano et al.(2019)]{romano} Romano D., Calura F., D\' Ercole A., Few C. G., 2019, A\&A, 630,A140
    
    \bibitem[Romanowsky \& Fall(2012)]{rf12} Romanowsky A. J., Fall S. M., 2012, ApJS, 203, 17
    
   \bibitem[Rong et al.(2017)]{rong} Rong Y., et al. 2017, MNRAS, 470, 4231
    
       \bibitem[Ruiz-Lara et al.(2018)]{ruizlara} Ruiz-Lara, T., et al. 2018, MNRAS, 478, 2034
    
    
    
    \bibitem[Saintonge et al.(2011)]{saintonge} Saintonge A., et al., 2011, MNRAS, 415, 32
    
    \bibitem[S\'anchez Almeida \& Filho(2019)]{sanchezalmeida} S\'anchez Almeida, J. \& Filho, M. 2019, RNAAS, 3 191 
    
            \bibitem[Sandage \& Binggeli(1984)]{sandage} Sandage \& Binggeli, 1984, AJ, 89, 919
    
        \bibitem[Sawala et al.(2016)]{apostle1} Sawala, T., Frenk, C. S., Fattahi, A., et al. 2016, MNRAS, 457, 1931

    \bibitem[Sancisi et al.(2008)]{renzo} Sancisi R., Fraternali, F., Oosterloo, T., van der Hulst, T. A\&ARv., 15, 189
    
    \bibitem[Sengupta et al.(2019)]{sengupta} Sengupta, C. et al. 2019, MNRAS, 488, 3222
    
    \bibitem[Spekkens \& Karunakaran(2018)]{spekkens} Spekkens K. \& Karunakaran A. 2018, ApJ, 855, 28
    
    \bibitem[Starkenburg et al.(2019)]{starkenburg} Starkenburg, T. K., Sales, L. V., Genel, S., et al. 2019,ApJ, 878, 143
    
    \bibitem[Strickland \& Stevens(2000)]{strickland} Strickland, D. K., \& Stevens, I. R. 2000, MNRAS, 314, 511
    
   
    
        \bibitem[Swaters(1999)]{swaters} Swaters, R. A. 1999, PhD thesis, University of Groningen
    
    
    
    \bibitem[Tamburro et al.(2009)]{tamburro} Tamburro, D. et al. 2009, AJ, 137, 4424. 
    
    \bibitem[Taylor(2005)]{topcat} Taylor, M. B. 2005, Astronomical Data Analysis Software and Systems XIV, 347, 29
    
    \bibitem[Teich et al.(2016)]{teich} Teich, Y. G., McNichols, A. T., Nims, E., et al. 2016, ApJ, 832, 85
    
    \bibitem[Thornton et al.(1988)]{thornton} Thornton K., Gaudlitz M., Janka H.-T., Steinmetz M., 1998, ApJ,500, 95
    
    \bibitem[Toloba et al.(2018)]{toloba} Toloba E., et al., 2018, ApJL, 856, L3
 
 \bibitem[Tremmel et al.(2019)]{tremmel} Tremmel, M. et al. 2019, submitted to MNRAS, arXiv:1908.05684
 
   \bibitem[Trujillo et al.(2017)]{trujillo} Trujillo, I., Rom\'an, J., Filho, M., \& S\'anchez Almeida, J. 2017, ApJ, 836, 191
    
    \bibitem[Utomo et al.(2019)]{utomo} Utomo, D., Blitz, L., \& Falgarone, E. 2019, ApJ, 871, 17
    
    
    
  \bibitem[van der Burg et al.(2016)]{vanderburg} van der Burg, R. F. J., Muzzin, A., \& Hoekstra, H. 2016, A\&A, 590, A20
   
   \bibitem[van der Hulst et al.(1992)]{gipsy1} van der Hulst J. M., Terlouw J. P., Begeman K. G., Zwitser W., Roelfsema P. R., 1992, in Worrall D. M.,  Biemesderfer C., Barnes J., eds, Astronomical Society of the  Pacific Conference Series Vol. 25, Astronomical Data Analysis Software and Systems I. p. 131

\bibitem[van der Kruit(1988)]{vanderkruit} van der Kruit P. C., 1988, A\&A, 192, 117
   
   \bibitem[van Dokkum et al.(2016)]{vandokkum16} van Dokkum, P., Abraham, R., Brodie, J., et al. 2016, ApJ, 828, L6

   \bibitem[van Dokkum et al.(2015)]{vandokkum} van Dokkum, P. G., Abraham, R., Merritt, A., et al. 2015, ApJL, 798,  L45




\bibitem[Venhola et al.(2017)]{Aku} Venhola, A., Peletier, R.F., et al. 2017, A\&A, 608, A142

\bibitem[Verheijen(1997)]{marc} Verheijen M. A. W. 1997, PhD thesis, University of Groningen


\bibitem[Vogelaar \& Terlouw(2001)]{gipsy2} Vogelaar M. G. R., Terlouw J. P., 2001, in Harnden Jr. F. R.,Primini F. A., Payne H. E., eds, Astronomical Society of the Pacific Conference Series Vol. 238, Astronomical Data Analysis Software and Systems X. p. 358

\bibitem[Vogelsberger et al.(2013)]{vogelsberger} Vogelsberger, M., Genel, S., Sijacki, D., et al. 2013,MNRAS, 436, 3031

\bibitem[Wang et al.(2015)]{nihao} Wang, L., et al. 2015, MNRAS, 454, 83

\bibitem[Yang et al.(2020)]{yang2020} Yang, D., Yu, H. B., An, H. 2020, arXiv:2002.02102

\bibitem[York et al.(2000)]{york} York D. G., et al., 2000, AJ, 120, 1579

\bibitem[Yun et al.(1994)]{yun} Yun, M. S., Ho, P. T. P., \& Lo, K. Y. 1994, Nature, 372, 530


\bibitem[Zwaan et al.(1995)]{zwaan95} Zwaan, M., van der Hulst, J., de Blok, W. \& McGaugh, S.,1995, MNRAS, 273, L35




\end{thebibliography}



\newpage
\appendix
\section{WSRT vs. VLA data}
\label{app:extra}

As mentioned in the main text, two galaxies in our sample, AGC~122966 and AGC~334315, have both VLA and WSRT data. Given the spatial resolution and $S/N$ of the data, we adopt as our `fiducial' data the WSRT observations for AGC~122966 and the VLA observations for AGC~334315. Here we delve into the reasons in which we based these choices, and we show the complementary data: the VLA data for AGC~122966 as well as the WSRT data for AGC~334315. We also make the comparison between the kinematic and geometrical parameters of the different data.

\begin{figure*}
    \centering
    \includegraphics[scale=0.5]{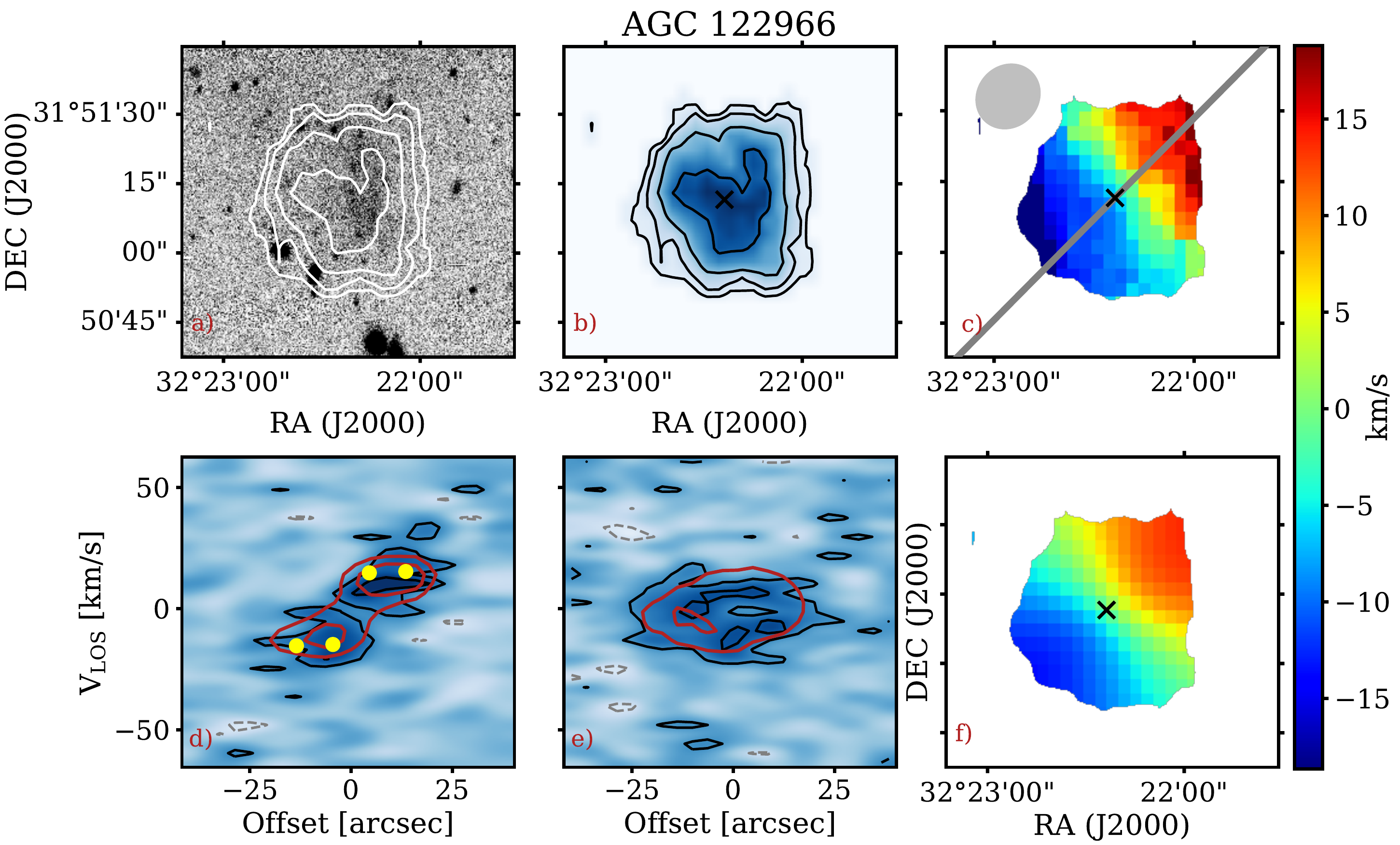}
    \caption{VLA data and kinematic models for AGC~122966. Panels and symbols as in Figure~\ref{fig:114905}. The H\,{\sc i} contours are at 0.5, 1, 2 and 4$\times$10$^{20}$ atoms~cm$^{-2}$.}
    \label{fig:122966VLA}
\end{figure*}

The VLA data cube of AGC~122966 has a lower $S/N$ than the WSRT one. Moreover, the emission is less extended, and we have to oversample by a factor 2 in order to fit the galaxy with $\mathrm{^{3D}}$Barolo. Figure~\ref{fig:122966VLA} shows the stellar image, H\,{\sc i} map, and observed and modelled velocity fields and PV diagrams. From inspecting the major-axis PV, we can see that the  emission connecting approaching and receding sides is not detected. More importantly, it looks likely that there is emission missing at the end of the approaching and residing sides too, which would significantly affect the recovered value of the rotation velocity. In fact, while for the WSRT cube $\mathrm{^{3D}}$Barolo finds a projected velocity of 26~km~s$^{-1}$, it finds 16~km~s$^{-1}$ for the VLA cube. Given all this, we decided to use the WSRT data for this galaxy (Figure~\ref{fig:122966}).

Despite the problems deriving the kinematics, we can still use the VLA data to independently estimate the inclination of the galaxy as described in the main text. Our method finds an inclination of 44$^\circ \pm 5^\circ$ using the WSRT H\,{\sc i} map and of 40$^\circ \pm 5^\circ$ with the VLA data, meaning that the inclination estimates from the different data cubes are consistent with each other. The VLA data of AGC~122966 is not only useful to confirm our inclination measurements, but also because the morphology of the galaxy can be better appreciated without the elongated beam of the WSRT observations (Figure~\ref{fig:122966}). Finally, the models for the different cubes have the same systemic velocity and physical center.

\begin{figure*}
    \centering
    \includegraphics[scale=0.5]{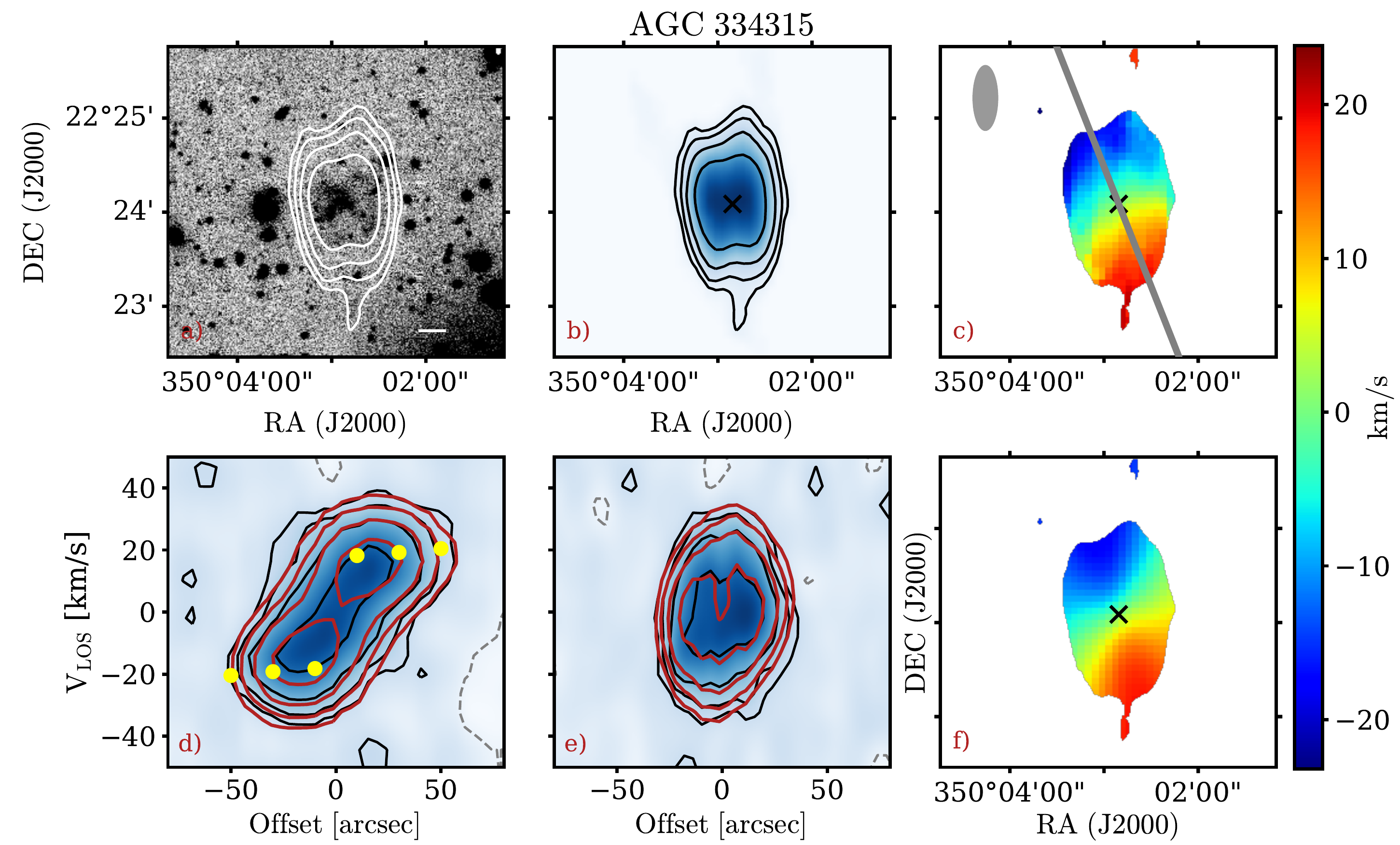}
    \caption{WSRT data and kinematic models for AGC~334315. Panels and symbols as in Figure~\ref{fig:114905}. The H\,{\sc i} contours are at 0.35, 0.7, 1.4 and 2.8$\times$10$^{20}$ atoms~cm$^{-2}$.}
    \label{fig:334315WSRT}
\end{figure*}

On the other hand, for AGC~334315 the VLA data (Figure~\ref{fig:334315}), have better quality (although a factor 2 less extended) than the WSRT data, shown in Figure~\ref{fig:334315WSRT}: the spatial and spectral resolution are better, and the beam is more circular. 
With this, one can more clearly appreciate the H\,{\sc i} structure as well as the intrinsic shape and velocity field of the galaxy. Apart from this, the parameters found for the two different data cubes are perfectly compatible with each other: we find an inclination of 45$^\circ \pm 5^\circ$ for the VLA map and of 52$^\circ \pm 5^\circ$ with the WSRT one, in agreement within the reported uncertainties. The final circular speed is also the same within the uncertainties: 25~km~s$^{-1}$ at 8.5~kpc for the VLA data, and 26~km~s$^{-1}$ at 18~kpc for the WSRT observations. The models for the different cubes share also the same systemic velocity and physical center.

{}

{}

\section{Comparison samples}
\label{app:comparisonsamples}
\subsection{The BTFR}
Figure~\ref{fig:BTFRall} shows the circular speed--baryonic mass plane, with galaxies scattering around the BTFR. The main result from the figure is that the region between our extreme UDGs and galaxies following the extrapolation of the SPARC BTFR at low masses, is populated by other dwarf galaxies. When discussing specific galaxies which are particularly interesting given their position with respect to the SPARC BTFR, we provide their coordinates on the $V_{\rm circ}-M_{\rm bar}$ plane.

\subsubsection{SPARC}

Within this sample (cyan circles in Figure~\ref{fig:BTFRall}), two galaxies deviate significantly from the SPARC BTFR: UGC~7125 (65~km~s$^{-1}$, 10$^{9.86}$~M$_\odot$) and UGC~9992 (34~km~s$^{-1}$, 10$^{8.77}$~M$_\odot$).

UGC~7125 has an inclination of 90$^\circ$ and $Q = 1$, so its position seems robust, even though, as \citet{lelli2016} mention, the galaxy has a relatively large uncertainty on its mass due to a high distance-correction for Virgocentric infall. 

UGC~9992 has an inclination of 30$^\circ \pm 10^\circ$ and $Q = 2$. Its PV diagram \citep{swaters} is regular and traces the flat part of the rotation curve. To bring this galaxy back to the BTFR an inclination of $\sim$~17$^\circ$ would be needed. For this hypothetical inclination to be correct, an error of 13$^\circ$ in the measured inclination is needed, a bit larger than the quoted uncertainty of 10$^\circ$ but within 1.5$\upsigma$. It is worth mentioning that two other galaxies with $i = 30^\circ \pm 10^\circ$ (F571-V1 and UGC~7261) lie much closer to the BTFR, as well as, for instance, the galaxies UGC~11914, UGC~6930 and UGC~10310, with $i = 31^\circ \pm 5^\circ, 32^\circ \pm 5^\circ$ and 34$^\circ \pm 6^\circ$, respectively.

\subsubsection{LITTLE THINGS}

Most LITTLE THINGS galaxies lie around the extrapolation of the SPARC BTFR, even when some of their rotation curves have \textit{not} clearly reached their flat part (see \citealt{iorio}). DDO~50 (also known as UGC~4305 or Ho II) is an outlier, in a position very close to our UDGs (39~km~s$^{-1}$, 10$^{8.95}$~M$_\odot$). Its rotation curve has clearly reached the flat part but the inclination of the galaxy is relatively low, 30$^\circ$. Different values have been proposed, ranging from 30$^\circ$ to 50$^\circ$ (see \citealt{kyle2016, iorio} and references therein), so the exact value of its circular speed remains somewhat uncertain. However, an inclination of $\sim$~18$^\circ$ is needed for the galaxy to lie directly on the canonical BTFR, which is outside the wide range proposed in the literature. 

IC~1613 (also known as DDO~8, not shown in Figure~\ref{fig:BTFRall}), is another well known candidate to be an outlier from the BTFR with a baryonic mass of 10$^{7.9}$~M$_\odot$ and rotation velocity of about 20~km~s$^{-1}$ \citep{oh15,kyle2016}. It is part of the LITTLE THINGS galaxies studied in \citet{oh15}, but was excluded in the work of \citet{iorio}. While the galaxy is potentially interesting, there may be issues with its inclination and equilibrium state, as mentioned in \citet{read}, Because of this, as in \citet{iorio}, we do not consider this galaxy in Figure~\ref{fig:BTFRall}.

In the sample from \citet{iorio}, DDO~101 (59~km~s$^{-1}$, 10$^{7.94}$~M$_\odot$) stands out but for being an outlier that rotates too fast for its baryonic mass. However, \citet{read} have demonstrated that this discrepancy is largely mitigated if the galaxy is farther away than the distance used in \citet{iorio}, so it is possible that the uncertainties in distance (and thus in baryonic mass) have been underestimated for this galaxy.

\subsubsection{SHIELD}
Due to the very low spatial resolution data the analysis of this sample was very challenging, as discussed in \citet{shield}. Five out of twelve galaxies in the sample have rotation velocities estimated from fitting tilted-ring models to the observed low-spatial resolution velocity fields, and at least three of the resulting rotation curves have no indication of a flattening. The remaining velocities were derived from the visual inspection of different PV slices (see \citealt{shield} for details). In addition to this, the asymmetric drift correction was not applied to the galaxies, and the inclination comes from optical data rather than from the H\,{\sc i} itself. Despite these differences with other analyses, the twelve galaxies lie around the extrapolation of the canonical BTFR.

\subsubsection{Edge-on HUDs}
\citet{he} selected a sample of nearly edge-on HUDs that were not originally selected in the catalogue of \citetalias{leisman} due to selection effects (given their high inclination the galaxies have apparent higher surface brightness). As the optical morphology indicates that the galaxies are edge-on, corrections for inclination are negligible. Their optical images, shown in \citet{he}, suggest that their stellar structure is not exactly the same as in our sample, as theirs look more regular and thinner, but it is hard to unequivocally judge this as our sample is not edge-on. However, the stellar component of our sample is both more extended (by a factor about 2.5) and of lower surface brightness.

Most of these galaxies (magenta diamonds in Figure~\ref{fig:BTFRall}) show velocities nearly compatible with the SPARC BTFR. The clear outlier is AGC~202262, that shows the narrowest velocity width, with a rotation of about 30~km~s$^{-1}$. An inclination of $\sim$~40$^\circ$, totally incompatible with the optical image (but see the caveats on the optical-gas misalignment reported in \citealt{starkenburg} or \citealt{gault}), would be required to put the galaxy on the canonical BTFR. 

\subsubsection{UGC~2162}

UGC~2162 is an UDG which has been studied using $\mathrm{^{3D}}$Barolo in resolved GMRT observations \citep{sengupta}. 
Two caveats exist regarding this galaxy. First, \citet{sengupta} mention that it has large uncertainties in its baryonic mass (probably because of the uncertainty associated with its distance; the galaxy is much closer than those in our sample), and second, its rotation curve does not show signs of flattening (the emission is less extended than in our data), although it is presumably close to reaching the flat part (see figure~2 in \citealt{kyle2015}).

While these caveats should be taken into account, the position of UGC~2162\footnote{UGC~2162 has no reported uncertainties in rotation velocity nor H\,{\sc i} mass by \citet{sengupta}, so we assume an uncertainty of 7~km~s$^{-1}$ for the velocity (the spectral resolution of the data, which is also larger than the difference in the velocity obtained from the best-fit model of $\mathrm{^{3D}}$Barolo and the global H\,{\sc i} profile) and a typical value of 20 per cent in the H\,{\sc i} mass.} in Figure~\ref{fig:BTFRall} (red hexagon) seems in line with our results. Note also that UGC~2162 is less `extreme' than our UDGs: it is less massive, smaller and slightly brighter. 
Assuming that the amplitude of the rotation curve does not increase significantly beyond the outermost measured radius, the galaxy would need an inclination of $\approx$~39$^\circ$ to be on the SPARC BTFR, far from the measured inclination of 55$^\circ$ in \citet{sengupta}. 

\subsection{Disc scale length and central surface brightness values}

In Figure~\ref{fig:VD_BTFR} we study the deviation from the BTFR as a function of the disc scale length; the central surface brightness of the galaxies was also inspected. We have obtained these structural parameters from the following sources in the literature. For the SPARC galaxies, scale lengths are taken directly from \citet{sparc} and surface brightnesses from \citet{marc,deblok} and \citet{swaters}. For LITTLE THINGS, scale lengths come from \citet{readAD} and surface brightnesses from \citet{hunterLT}. Regarding SHIELD, size and surface brightness come from \citet{teich} and \citet{haurberg}, respectively. \citet{he} provides both size and surface brightness for their sample, and the values used for UGC~2162 come from \citet{trujillo}.


\bsp	
\label{lastpage}
\end{document}